\documentclass[aps,prl,reprint,lengthcheck,longbibliography,superscriptaddress]{revtex4-2}
\usepackage[T1]{fontenc}
\usepackage{amsmath}
\usepackage{amsfonts}
\usepackage{amssymb}
\usepackage{graphicx}
\usepackage{color}
\usepackage{physics}
\usepackage[colorlinks,allcolors=blue]{hyperref}

\newcommand{\al}[1]{{\color{black} #1}}

\begin{document}
	\newcommand{\FTMC}{\affiliation{Departamento de F\'{\i}sica Te\'orica de la Materia Condensada,
			Universidad Aut\'onoma de Madrid, E-28049 Madrid, Spain}}
	\newcommand{\IFIMAC}{\affiliation{Condensed Matter Physics Center (IFIMAC),
		Universidad Aut\'onoma de Madrid, E-28049 Madrid, Spain}}
	\newcommand{\RPTU}{\affiliation{Department of Physics and Research Center OPTIMAS, 
			Rheinland-Pfälzische Technische Universität Kaiserslautern-Landau, 67663 Kaiserslautern, Germany}}

	\title{Squeezing and quantum control of antiferromagnetic magnon pseudospin}	
	\author{Anna-Luisa E. R\"omling}
	\email{anna-luisa.romling@uam.es}
	\FTMC \IFIMAC

	\author{Johannes Feist}
	\FTMC \IFIMAC

	\author{Francisco J. Garc\'{\i}a-Vidal}
	\FTMC \IFIMAC

	\author{Akashdeep Kamra}
	\RPTU \FTMC \IFIMAC

	\begin{abstract}
		Antiferromagnets have been shown to harbor strong magnon squeezing in equilibrium, making them a potential resource for quantum correlations and entanglement. Recent experiments have also found them to host coherently coupled magnonic excitations forming a magnon pseudospin, in analogy to electronic spin. Here, we delineate the quantum properties of antiferromagnetic magnon pseudospin by accounting for spin non-conserving interactions and going beyond the rotating wave approximation. Employing concrete examples of nickel oxide and hematite, we find strong squeezing of the magnon pseudospin highlighting its important role in determining the eigenmode quantum properties. Via ground state quantum fluctuations engineering, this pseudospin squeezing enables an enhancement and control of coupling between the magnonic modes and other excitations. Finally, we evaluate the quantum superpositions that comprise a squeezed pseudospin ground state and delineate a qubit spectroscopy protocol to detect them. Our results are applicable to any system of coupled bosons and thus introduce quantum fluctuations engineering of a general bosonic pseudospin. 
	\end{abstract}
	\maketitle
	
	\textit{Introduction.---}As per the Heisenberg uncertainty~\cite{heisenberg1927} relation, complementary physical observables cannot be measured precisely at the same time. Squeezed states harbor reduced quantum fluctuations or uncertainty in a physical observable at the expense of an increased noise in its complementary counterpart~\cite{gerry2005, walls2008,schnabel2017}. For light, such states have thus been exploited for measurements with a sensitivity beyond what is allowed by the quantum ground state and coherent states as produced by lasers~\cite{abbott2009, schnabel2017,grote2013, aasi2013}. Contained implicitly in this engineering of quantum fluctuations is entanglement, which has been utilized for teleportation~\cite{milburn1999,bennett1993,hoke2023}. The generality of the concept has enabled the realization of squeezed states with one boson mode~\cite{slusher1985,wu1986,vahlbruch2016}, two boson modes~\cite{steinlechner2013,ast2016}, and atomic ensembles of spin~\cite{hald1999,kuzmich1997,hammerer2010,bohnet2014,bohnet2016,hosten2016,hosten2016a,luo2017,zou2018,bao2020,eckner2023}, with each platform harboring unique phenomena and advantages. However, the squeezed states in these examples are nonequilibrium in nature and decay rapidly as the generating mechanism is switched off.
	
	Ordered magnets and their excitations – magnons – have recently been shown to harbor squeezed states and excitations in equilibrium, thereby uncovering fresh advantages and phenomena~\cite{kamra2016,kamra2019,kamra2020,yuan2020,zou2020,wuhrer2022,azimimousolou2021,wuhrer2024}. Magnons have already proven promising for an information transport and processing paradigm harnessing the unique potential of spin and bosonic excitations, with several phenomena and devices demonstrated~\cite{lachance-quirion2019,kamra2020, yan2022, awschalom2021,shim2020, psaroudaki2021}. Antiferromagnetic magnons are particularly exciting due to their typically high frequencies and equilibrium squeezing larger than what has been feasible with light~\cite{kamra2020, yuan2020}. Furthermore, they come in pairs of spin-up and spin-down modes, much like the spin of an electron. Coherently coupled spin-up and spin-down modes enable a new degree of freedom – magnon pseudospin – capable of capitalizing on the positive features of fermionic electrons, bosonic magnons, and squeezing~\cite{kamra2020a,wimmer2020}. The recent observation of the magnon Hanle effect~\cite{wimmer2020} has demonstrated experimental control over the magnon pseudospin and uncovered a wide range of fresh opportunities\cite{kleinherbers2024,tang2024,ross2020,guckelhorn2023,sheng2025}.
	
	Inspired by its usefulness in several newly discovered spin transport phenomena, here, we uncover some quantum properties of the antiferromagnetic magnon pseudospin focusing on its squeezing and the concomitant eigenmode quantum superpositions. We demonstrate that its mathematical similarity to spin offers an equilibrium realization of spin squeezing via unequal quantum fluctuations in orthogonal pseudospin components. This pseudospin squeezing can be conveniently controlled via typical spintronic controls, such as an applied magnetic field. At the same time, the underlying bosonic nature of the pseudospin enables fresh coupling enhancement and control opportunities akin to similar effects observed recently in, for example, trapped ion systems~\cite{burd2021}. The quantum superpositions of a squeezed state that underlie these phenomena are then evaluated for the magnon pseudospin and a protocol for detecting them via a qubit is theoretically demonstrated. Employing experimentally measured parameters, we quantify the pseudospin squeezing in nickel oxide and hematite, and note the recently discovered van der Waals magnets to naturally embody our studied coupled magnon pseudospin-qubit system~\cite{klein2024, melendez2025}.
	
	\textit{Antiferromagnetic magnon pseudospin.---}We consider a bipartite antiferromagnetic insulator (AFM)~\cite{keffer1952,kittel1987} with exchange as the dominant interaction. In the N\'eel ordered classical ground state, the two spin sublattices are fully and oppositely polarized. We denote sublattice $A$ ($B$) as the sublattice polarized along the negative (positive)  $z$-direction. Excitations on sublattice $A$ ($B$) are delocalized spin-flips with spin $+1$ ($-1$). However, strong exchange coupling between the sublattices induces correlations and intrinsic squeezing, quantified by a parameter $r$, in the quantum fluctuations of the sublattice spins~\cite{kamra2019}. Because of this, excitations of the AFM are spin-up and spin-down magnonic modes perturbing both sublattices as a superposition of spin-flip states. We consider here the spatially uniform modes (wave vector $\boldsymbol{k} = \boldsymbol{0}$), and represent spin-up and spin-down magnonic modes by the bosonic annihilation operators $\hat{\alpha}$ and $\hat{\beta}$.
	
	Accounting for the spin non-conserving (SNC) interactions, coherent magnon-magnon coupling between the modes $\hat{\alpha}$ and $\hat{\beta}$ is obtained, as derived in the Supplemental Material (SM)~\cite{SM}. Within the rotating wave approximation (RWA), an assumption we later relax, the magnonic excitations are described by the following Hamiltonian ($\hbar=1$)~\cite{kamra2020a}
	\begin{figure}
		\centering
		\includegraphics[width = 0.6\columnwidth]{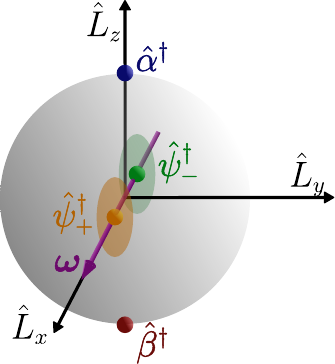}
		\caption{Representation of the eigenmodes $\hat{\psi}_\pm^\dagger$ on a unit sphere with uncertainty area of the non-commuting pseudospin components $\hat{L}_y$ and $\hat{L}_z$. The sphere poles are the pure spin-up and spin-down magnons, represented by creation operators $\hat{\alpha}^\dagger$ and $\hat{\beta}^\dagger$. The considered pseudofield $\boldsymbol{\omega}$ is along the $\hat{L}_x$-axis and characterizes spin-0 modes. \label{fig:Fig1}}
	\end{figure}
	\begin{equation}
		\hat{\mathcal{H}}_\mathrm{Hyb}=\omega\left(\hat{\alpha}^{\dagger}\hat{\alpha}+\hat{\beta}^{\dagger}\hat{\beta}\right)-D_{r}\left(\hat{\alpha}\hat{\beta}^{\dagger}+\hat{\alpha}^{\dagger}\hat{\beta}\right), \label{eq:Hamiltonian1}
	\end{equation}
	where $\omega$ denotes the frequency of uncoupled $\hat{\alpha}$, $\hat{\beta}$ modes and $D_r$ quantifies their mutual coherent coupling, which has its physical origin in the SNC interactions. Without loss of generality, we assume $D_r$ to be real~\cite{SM, azimi-mousolou2024, kamra2020a} noting that its magnitude increases exponentially with the intrinsic squeezing parameter $r$~\cite{SM}. The eigenmodes of Eq.~\eqref{eq:Hamiltonian1} stem from hybridization between magnon modes $\hat{\alpha}$ and $\hat{\beta}$, and can be conveniently obtained using the antiferromagnetic magnon pseudospin~\cite{kamra2020a} 
	\begin{equation}
		\begin{split}
			\hat{L}_{0}=\frac{1}{2}\left(\hat{\alpha}^{\dagger}\hat{\alpha}+\hat{\beta}^{\dagger}\hat{\beta}\right),~~
		\end{split}
		\begin{split}
		\hat{\boldsymbol{L}}=\frac{1}{2}\left(\begin{array}{c}
			\hat{\alpha}\hat{\beta}^{\dagger}+\hat{\alpha}^{\dagger}\hat{\beta}\\
			i\hat{\alpha}\hat{\beta}^{\dagger}-i\hat{\alpha}^{\dagger}\hat{\beta}\\
			\hat{\alpha}^{\dagger}\hat{\alpha}-\hat{\beta}^{\dagger}\hat{\beta}
		\end{array}\right),
		\end{split}
		\label{eq:pseudospin}
	\end{equation}
	with commutation relations for angular momentum $[\hat{L}_j,\hat{L}_k] = i\epsilon_{jkl}\hat{L}_l$, $j,k,l\in \{x,y,z\}$. With Eq.~\eqref{eq:pseudospin}, $\hat{\mathcal{H}}_\mathrm{Hyb}$ [Eq.~\eqref{eq:Hamiltonian1}] transforms into $\hat{\mathcal{H}}_\mathrm{Hyb} = 2\omega\hat{L}_{0} - \boldsymbol{\omega}_0\cdot\hat{\boldsymbol{L}}$. The eigenmmodes are conveniently obtained via the points where the pseudofield $\boldsymbol{\omega}_0 = 2D_r \hat{\boldsymbol{e}}_x$ intersects with a unit sphere [Fig.~\ref{fig:Fig1}]~\cite{kamra2020a}, in direct analogy to how electronic spin eigenstates are determined via intersection of the experienced magnetic field with the Bloch sphere.
	 
	\textit{Magnon pseudospin squeezing.---}In order to account for quantum fluctuations, we now go beyond the RWA and explicitly account for the counter rotating terms. The total magnonic Hamiltonian then reads~\cite{rezende2019, SM}
	 \begin{align}
		\hat{\mathcal{H}}_\mathrm{AFM}=&\omega\left(\hat{\alpha}^{\dagger}\hat{\alpha}+\hat{\beta}^{\dagger}\hat{\beta}\right)-D_{r}\left(\hat{\alpha}\hat{\beta}^{\dagger}+\hat{\alpha}^{\dagger}\hat{\beta}\right)\nonumber\\
		&+D_{s}\left(\hat{\alpha}^{2}+\hat{\alpha}^{\dagger2}+\hat{\beta}^{2}+\hat{\beta}^{\dagger2}\right),\label{eq:Hamiltonian2}
	\end{align} 
	with coupling strength $D_s$ stemming from SNC interaction~\cite{SM}. On diagonalization, the eigenmodes of $\hat{\mathcal{H}}_\mathrm{AFM}$ [Eq.~\eqref{eq:Hamiltonian2}] are annihilated by $\hat{\psi}_\pm$ with the corresponding Fock eigenstates $\ket{m_+, n_-}$. As detailed in the SM~\cite{SM}, $\ket{m_+, n_-}$ are related to the magnon Fock states $\ket{m_\alpha, n_\beta}$ -- the eigenstates of $\omega(\hat{\alpha}^\dagger\hat{\alpha}+\hat{\beta}^\dagger\hat{\beta})$ --via a rotation of the pseudofield and two squeeze operations $\hat{S}_\pm\!(r_\pm)$:
	\begin{equation}
		\ket{m_+,n_-} = \hat{S}_+\!(r_+)\hat{S}_-\!(r_-)\hat{R}_y\!\left(\frac{\pi}{2}\right)\ket{m_\alpha,n_\beta}, \label{eq:eigenstates}
	\end{equation}
	with $\hat{R}_y\!(\phi)=\exp\bigl(-i\phi\hat{L}_y\bigr)$, where $\phi$ is defined as the angle between the $z$-axis and pseudofield, $\hat{S}_+\!(x) = \exp\bigl(\frac{x}{2}\hat{R}_y\!(\frac{\pi}{2})\left[\hat{\alpha}^2- \hat{\alpha}^{\dagger2}\right]\hat{R}_y^\dagger\!(\frac{\pi}{2})\bigr)$, where $\hat{S}_-\!(x)$ is obtained from $\hat{S}_-(x)$ upon substitution $\hat{\alpha} \rightarrow \hat{\beta}$, and quadrature squeeze factors $r_\pm$
	\begin{align}
		\tanh&(r_\pm) = \frac{2|D_s|}{\omega\mp D_r}, \label{eq:oms} 
	\end{align}
	where the ground state stability requires and ensures $2|D_s| < (\omega - |D_r|)$ and $|D_r|<\omega$.
	
	We find that SNC interactions in $\hat{\mathcal{H}}_\mathrm{AFM}$ [Eq.~\eqref{eq:Hamiltonian2}] induce what we term magnon pseudospin squeezing, in system eigenstate $\ket{m_+,n_-}$ [Eq.~\eqref{eq:eigenstates}]. $\hat{L}_y$ and $\hat{L}_z$ are non-commuting $[ \hat{L}_y,\hat{L}_z] = i\hat{L}_x$ and therefore obey Heisenberg's uncertainty relation. Considering a special subset of eigenstates where $m\neq0$ and $n=0$, as a simpler example, we determine the following expectation value $\langle \hat{L}_x \rangle_{m0}$ and variances $\langle \Delta\hat{L}_{i}^{2}\rangle _{m0} = \langle \hat{L}_{i}^{2}\rangle _{m0} - \langle \hat{L}_{i}\rangle _{m0}^{2}$, $i\in y,z$
	\begin{align}
		\left\langle \hat{L}_x \right\rangle_{m0} &=\frac{m +\frac{1}{2}}{2}\cosh(2r_{+})-\frac{\cosh(2r_{-})}{4}, \label{eq:Lx}\\
		\left\langle \Delta\hat{L}_{y}^{2}\right\rangle _{m0}&=\frac{m+\frac{1}{2}}{4}\cosh\left(2r_{+}-2r_{-}\right)-\frac{1}{8},\label{eq:xi_y}
	\end{align}
	where $\langle \Delta\hat{L}_{z}^{2}\rangle _{m0}$ can be obtained from $\langle \Delta\hat{L}_{y}^{2}\rangle _{m0}$ [Eq.~\ref{eq:xi_y}] upon substitution $-2r_- \rightarrow + 2r_-$.  We find that the area of uncertainty is given by $\langle \Delta\hat{L}_{y}^{2}\rangle_{m0}\langle \Delta\hat{L}_{z}^{2}\rangle_{m0}=\bigl|\langle \hat{L}_{x}\rangle _{m0}\bigr|^2/4 + \left(m^{2}+m\right)\left(\cosh4r_{-}-1\right)/8$ and therefore the state $\ket{m_+,0_-}$ is not minimum uncertainty since $\langle \Delta\hat{L}_{y}^{2}\rangle_{m0}\langle \Delta\hat{L}_{z}^{2}\rangle_{m0}>\bigl|\langle \hat{L}_{x}\rangle _{m0}\bigr|^2/4$. However, the variance in the $y$-component fulfils $\langle\Delta\hat{L}_{y}^{2}\rangle _{m0}<\bigl|\langle \hat{L}_{x}\rangle _{m0}\bigr|/2$~\cite{SM} and is hence squeezed.
	
\begin{figure}
		\centering
		\includegraphics[width = 0.9\columnwidth]{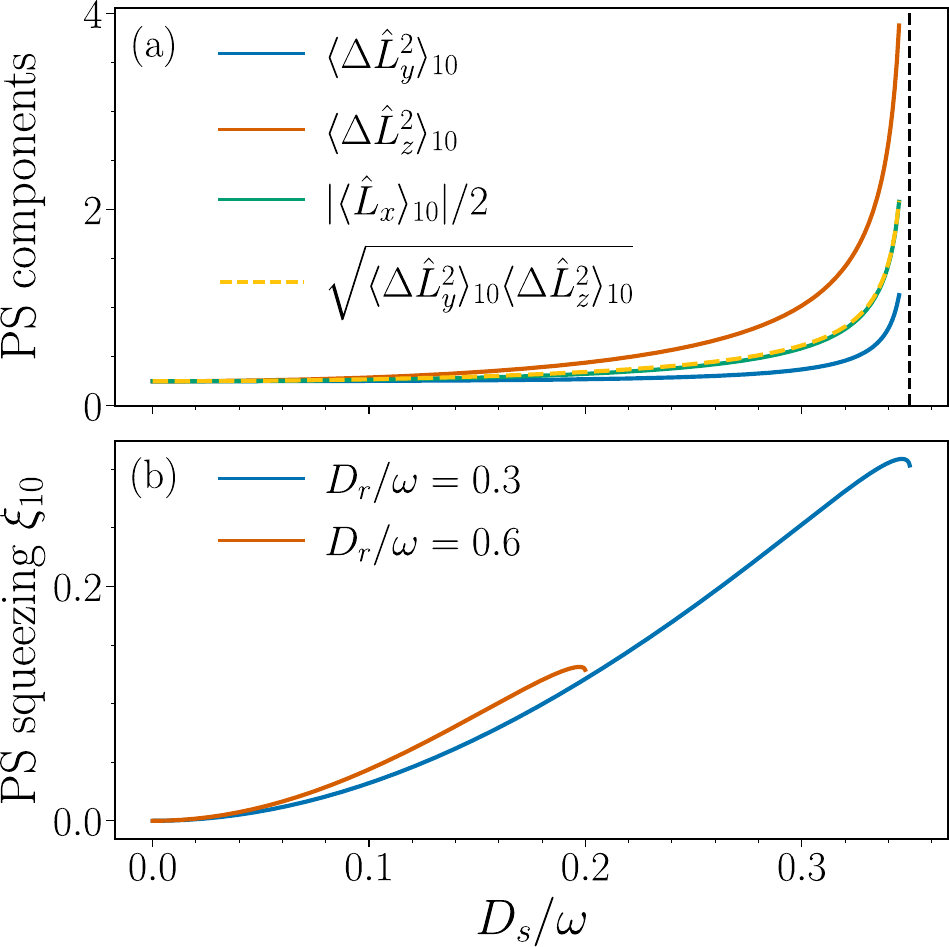}
		\caption{(a) Expectation value and variances of pseudospin (PS) components $\langle\hat{L}_x\rangle$, $\langle \Delta\hat{L}_y^2\rangle$ and $\langle \Delta\hat{L}_z^2\rangle$ as a function of $D_s$ with $D_r/\omega = 0.3$, $m=1$ and $n=0$. (b) Pseudospin squeezing $\xi_{10}$ as a function of $D_s$ for different values of $D_r$. The curves are limited by the ground state stability condition  $2|D_s| < (\omega - |D_r|)$. \label{fig:Fig2}}
	\end{figure}

	In order to quantify squeezing, we define the pseudospin squeezing factor $\xi_{m0}$ by the ellipticity of quantum fluctuations in the $\hat{L}_y$-$\hat{L}_z$ plane~\cite{gerry2005,SM}, which is explicitly given by  
	\begin{equation}
		\xi_{m0}=-\frac{1}{4}\ln(\frac{\left\langle \Delta\hat{L}_{y}^{2}\right\rangle _{m0}}{\left\langle \Delta\hat{L}_{z}^{2}\right\rangle _{m0}}). \label{eq:xi}
	\end{equation}
	In the schematics of Fig.~\ref{fig:Fig1}, we qualitatively sketch the eigenmodes $\hat{\psi}_\pm$ on the Bloch sphere with the corresponding squeezed uncertainty regions in the $\hat{L}_y$-$\hat{L}_z$ plane. In Fig.~\ref{fig:Fig2}(a) we plot expectation value and variances of pseudospin components $\langle\hat{L}_x\rangle_{10}$, $\langle \Delta\hat{L}_y^2\rangle_{10}$ and $\langle \Delta\hat{L}_z^2\rangle_{10}$ as a function of $D_s$, demonstrating that $\langle \Delta\hat{L}_y^2\rangle_{10} < |\langle\hat{L}_x\rangle_{10}|/2$. In Fig.~\ref{fig:Fig2}(b), we plot pseudospin squeezing $\xi_{10}$ [Eq.~\eqref{fig:Fig2}] as a function of $D_s$ for two values of $D_r$ showing that $\xi_{10}$ is a steadily growing function of $D_s$, i.e. $\xi_{10}=0$ if $D_s=0$. Therefore, $D_s$ is responsible for pseudospin squeezing. At the instability, where $2|D_s| = (\omega - |D_r|)$, pseudospin squeezing approaches $\xi_{10} \to r_-$~\cite{SM}.
	
	\textit{Coupling enhancement.---} Exponentially enhanced coupling between a squeezed bosonic mode and another quasiparticle has been demonstrated theoretically and experimentally in a wide range of systems~\cite{walls2008,gerry2005,schnabel2017}. Equilibrium squeezing inherits this property and enables enhanced coupling between, for example, itinerant electrons of normal metals (NM) and AFM magnons~\cite{kamra2019} when the interaction is mediated by sublattice spin ~\cite{kamra2019,zhang2016,kamra2017a,kamra2018}. We now employ this principle in demonstrating an additional enhancement and control of coupling between an antiferromagnetic magnonic mode and itinerant electrons, using them as an example, via the magnon pseudospin squeezing.
	
\begin{figure}
		\centering
		\includegraphics[width = 0.9\columnwidth]{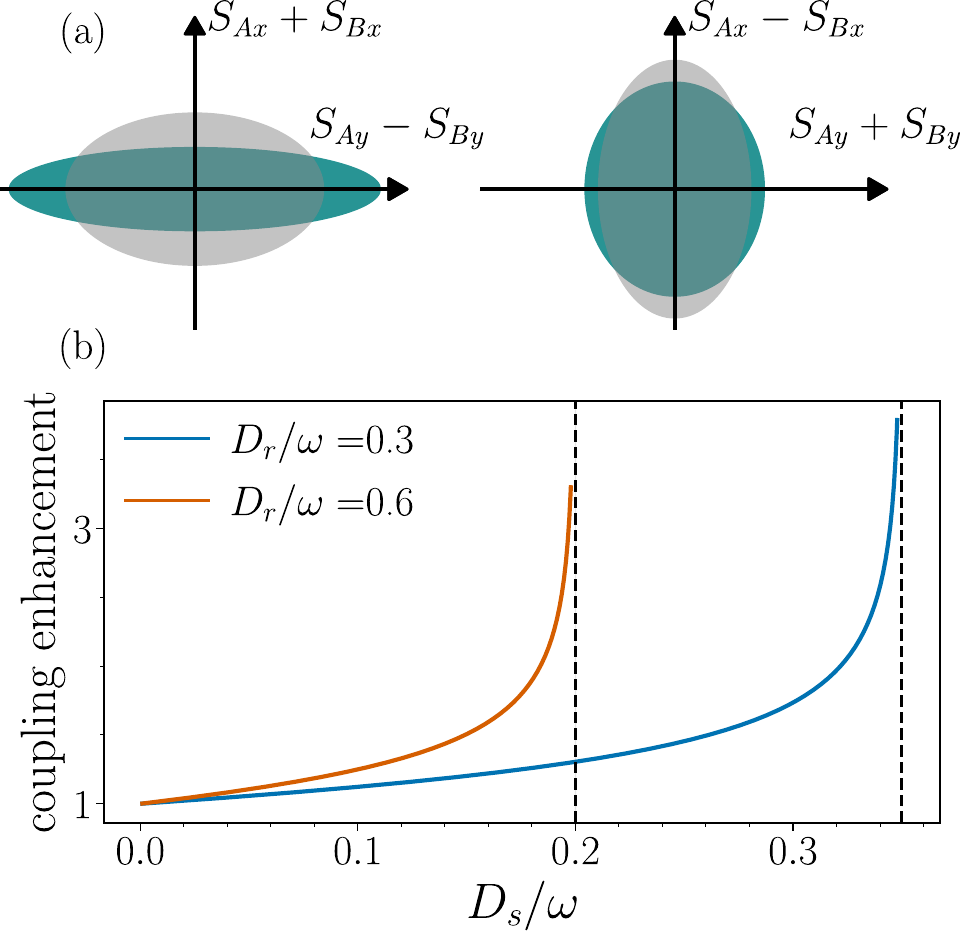}
		\caption{(a) Uncertainty region of the quadratures $\hat{X}_{1/2}$ and $\hat{Y}_{1/2}$ of sublattice spin operators $\hat{\boldsymbol{S}}_{A/B}$ for an AFM with $D_s\approx 0$ (grey) and AFM with SNC interactions (green). (b) Coupling enhancement $\cosh(r + r_+)/\cosh r$ of $\psi_+$ in the ground state $\ket{0_+, 0_-}$ for an uncompensated interface as a function of $D_s$, with intrinsic squeezing $r=1$.\label{fig:Fig3}}
	\end{figure}

	Let's denote sublattice spin operators $\hat{\boldsymbol{S}}_{A}$ and $\hat{\boldsymbol{S}}_{B}$ where $A$ ($B$) signifies the spin-down (up) sublattice and define the four quadrature operators $\hat{X}_{1/2}	=\bigl(\hat{S}_{Ax/y}+\hat{S}_{Bx/y}\bigr)/\sqrt{4NS}$ and $\hat{Y}_{1/2}	=\bigl(-\hat{S}_{Ay/x}+\hat{S}_{By/x}\bigr)/\sqrt{4NS}$. Here, $N$ denotes the number of sublattice sites and $S$ the spin of a single lattice site. In an AFM with negligible quadratic interactions ($D_s\approx 0$), the ground state fluctuations of quadrature operator $\hat{X}_{1/2}$ (sum of sublattice spins) is reduced in comparison to the variance of $\hat{Y}_{1/2}$ (difference of sublattice spins) because of intrinsic squeezing [see Fig~\ref{fig:Fig3}(a)]. The sublattice spins prefer to stay antiparallel, therefore forming correlations~\cite{kamra2019,kamra2020}. In the eigenmodes harboring magnon pseudospin squeezing evaluated above, we find that the variances of the quadrature operators $\hat{X}_{1/2}$ and $\hat{Y}_{1/2}$ in the ground state $\ket{0_+,0_-}$ [Eq.~\eqref{eq:eigenstates}] read
	\begin{align}
			\left\langle \Delta\hat{X}_{1/2}^{2}\right\rangle _{00}	=\frac{e^{-2\left(r+r_{\pm}\right)}}{4},
			\left\langle \Delta\hat{Y}_{1/2}^{2}\right\rangle _{00}	=\frac{e^{2\left(r+r_{\pm}\right)}}{4}.\label{eq:sublattice_fluctuations}
	\end{align} 
	In comparison with an AFM with $D_s\approx 0$, where $\bigl\langle \Delta\hat{X}_{1}^{2}\bigr\rangle _{00}=\bigl\langle \Delta\hat{X}_{2}^{2}\bigr\rangle _{00}$ and $\bigl\langle \Delta\hat{Y}_{1}^{2}\bigr\rangle _{00}=\bigl\langle \Delta\hat{Y}_{2}^{2}\bigr\rangle _{00}$~\cite{kamra2020}, we find that SNC interactions quadratic in magnon operators $\hat{\alpha}$ and $\hat{\beta}$ cause an asymmetry between the $x$ and $y$ components of spin operators, as illustrated in the schematics of Fig.~\ref{fig:Fig3}(a). This introduces a new way of engineering quantum fluctuations in AFMs. 
	
	Focusing on a concrete example of AFM interfaces with a normal metal, we consider interaction between electrons and the AFM modes via interfacial exchange coupling between sublattice spins and electron spin:~\cite{kamra2019, bender2012, zheng2017}
	\begin{align}
	\hat{\mathcal{H}}_{\text{NA}}	=&\sum_{\boldsymbol{q}_{1},\boldsymbol{q}_{2}} \hat{c}_{\boldsymbol{q}_{1}\downarrow}^{\dagger}\hat{c}_{\boldsymbol{q}_{2}\uparrow}\Big[\big(W_{\boldsymbol{q}_{1}\boldsymbol{q}_{2}}^{A}\cosh\!r + W_{\boldsymbol{q}_{1}\boldsymbol{q}_{2}}^{B} \sinh\!r\big) \hat{\alpha}\nonumber\\
	&+\big(W_{\boldsymbol{q}_{1}\boldsymbol{q}_{2}}^{A}\cosh\!r + W_{\boldsymbol{q}_{1}\boldsymbol{q}_{2}}^{A} \sinh\!r\big) \hat{\beta}^\dagger\Big]+\text{h.c.},\label{eq:H_NA1}
	\end{align}
	where $\hat{c}_{\boldsymbol{q}\sigma}$ denotes the annihilation operator of an electron with momentum $\boldsymbol{q}$ and spin $\sigma$ and $W_{\boldsymbol{q}_{1}\boldsymbol{q}_{2}}^{A/B}$ the scattering amplitude, which includes information about the interfacial coupling strength with sublattice $A$ and $B$ respectively. In the eigenbasis $\psi_{\pm}$, the scattering Hamiltonian $\hat{\mathcal{H}}_\mathrm{NA}$ [Eq.~\eqref{eq:H_NA1}] reads
	\begin{align}
		\hat{\mathcal{H}}_{\text{NA}}=\sum_{\boldsymbol{q}_{1},\boldsymbol{q}_{2}, \gamma=\pm}\hat{c}_{\boldsymbol{q}_{1}\downarrow}^{\dagger}\hat{c}_{\boldsymbol{q}_{2}\uparrow} \Big(&U_\gamma\psi_{\gamma} +V_\gamma\psi_{\gamma}^{\dagger}\Big) +\text{h.c.},
	\end{align}
	with amplitudes
	\begin{align}
		U_\pm&=W_{\boldsymbol{q}_{1}\boldsymbol{q}_{2}}^{A}\frac{\cosh(r\pm r_{\pm})}{\sqrt{2}}-W_{\boldsymbol{q}_{1}\boldsymbol{q}_{2}}^{B}\frac{\sinh\left(r\pm r_{\pm}\right)}{\sqrt{2}},\label{eq:amplitudes}
	\end{align} 
	where $V_+(V_-)$ can be obtained from  $U_+(U_-)$ upon exchange $W_{\boldsymbol{q}_{1}\boldsymbol{q}_{2}}^{A}\leftrightarrow W_{\boldsymbol{q}_{1}\boldsymbol{q}_{2}}^{B}$. We find that for a compensated interface ($W_{\boldsymbol{q}_{1}\boldsymbol{q}_{2}}^{A}=W_{\boldsymbol{q}_{1}\boldsymbol{q}_{2}}^{B}$) the scattering amplitudes [Eq.~\eqref{eq:amplitudes}] are exponentially suppressed by $e^{-\left(r+r_+\right)}$ and $e^{-\left(r-r_-\right)}$, whereas for an uncompensated interface with $W_{\boldsymbol{q}_{1}\boldsymbol{q}_{2}}^{A}=0$ and $W_{\boldsymbol{q}_{1}\boldsymbol{q}_{2}}^{B}\neq0$ the creation of the $\psi_{+}^{\dagger}$ mode is enhanced by $\cosh(r+r_+)$. This feature allows for mode selection and modification of scattering amplitudes by tuning the parameters $D_r$ and $D_s$. In Fig.~\ref{fig:Fig3}(b) we plot the coupling enhancement $\cosh(r+r_+)$ as a function of $D_s$ and normalized to the coupling enhancement of an AFM with $D_s\approx 0$ given by $\cosh r$. \\
	
\begin{figure}
		\centering
		\includegraphics[width = 0.9\columnwidth]{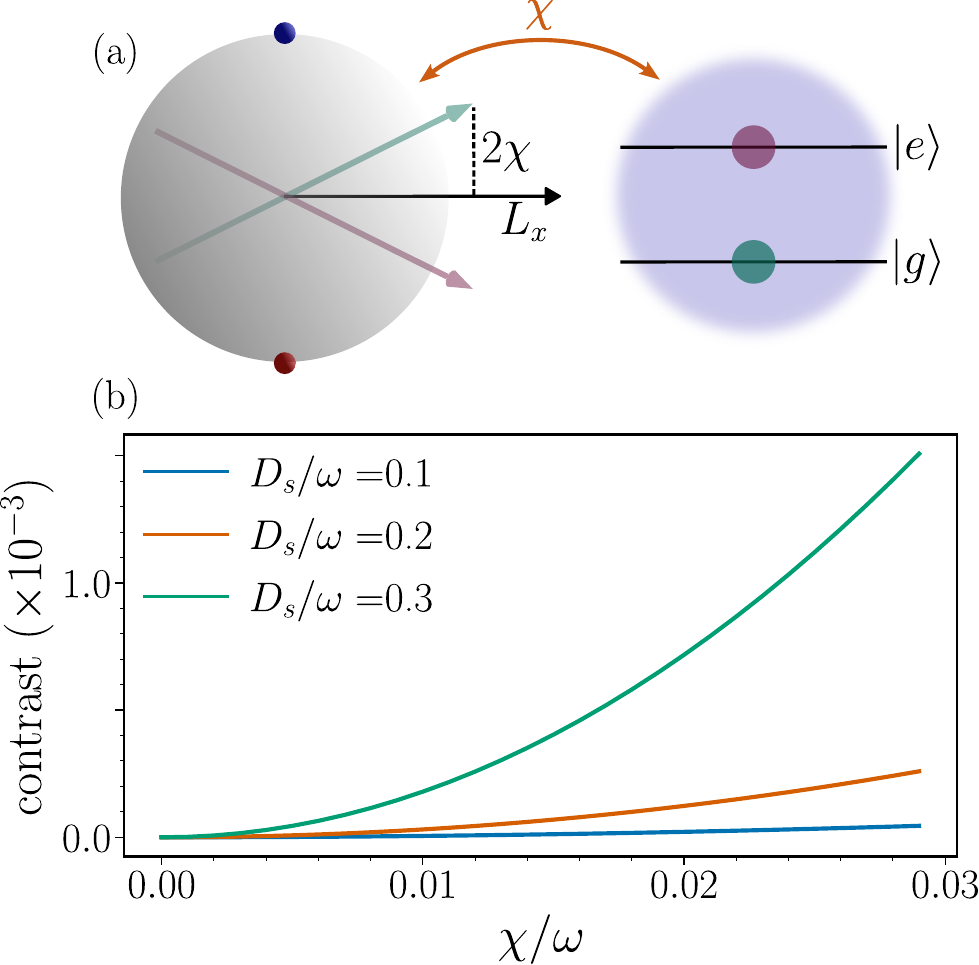}
		\caption{(a) Control of pseudofield via qubit in the ground state $\ket{g}$ (green) and excited state $\ket{e}$ (pink). (b) Contrast of the nontrivial peak in the small $\chi/D_r \ll1$ approximation as a function of $\chi$ with $D_r/\omega = 0.3$ for three values of $D_s$.\label{fig:Fig4}}
	\end{figure}

	\textit{Quantum control and superposition resolution via a qubit.---} A hallmark of squeezed states is that they are comprised by quantum superpositions that underlie nearly all their unique properties. Experimentally detecting these intrinsic superpositions should offer direct access to their quantum features. Recent theoretical proposals suggest a direct dispersive interaction between spin qubits and magnets for probing these superpositions in the ground state via qubit spectroscopy~\cite{romling2023,romling2024}. Here, we show that this interaction enables a qubit-state dependent pseudofield thereby enabling a quantum control of the AFM ground state and eigenmodes [see Fig.~\ref{fig:Fig4}(a)]. 
	
	We consider direct dispersive coupling between a spin qubit and AFM with a compensated interface, described by the following interaction Hamiltonian~\cite{romling2024} 
	\begin{equation}
		\hat{\mathcal{H}}_{\text{qcon}}=\chi\left(\hat{\alpha}^{\dagger}\hat{\alpha}-\hat{\beta}^{\dagger}\hat{\beta}\right)\hat{\sigma}_z=2\chi \hat{L}_z \hat{\sigma}_z, \label{eq:H_qcom}
	\end{equation}
	with the interaction strength $\chi$~\cite{SM}. The Hamiltonian representing the coupled magnet-qubit system reads $\hat{\mathcal{H}}_\mathrm{mq} = \hat{\mathcal{H}}_\mathrm{AFM} + \hat{\mathcal{H}}_\mathrm{q} + \hat{\mathcal{H}}_\mathrm{qcon}$, where $\hat{\mathcal{H}}_\mathrm{q}=\omega_{q}\hat{\sigma}_{z}/2$ stands for the spin qubit with level splitting $\omega_q$. To evaluate the effect of the qubit on the pseudospin state, we consider the reduced Hamiltonians $\hat{\mathcal{H}}_g = \bra{g}\hat{\mathcal{H}}_\mathrm{mq} \ket{g}$ and $\hat{\mathcal{H}}_e = \bra{e} \hat{\mathcal{H}}_\mathrm{mq} \ket{e}$, where $\ket{g}$ ($\ket{e}$) denotes the qubit ground (excited) state.
	
	The eigenstates of $\hat{\mathcal{H}}_{g/e}$, denoted by $\ket{m,n}_{g/e}$, are obtained from Eqs.~\eqref{eq:eigenstates} and \eqref{eq:oms} upon substitution $D_r \rightarrow \sqrt{D_r^2 + \chi^2}$ and $\pi/2 \rightarrow \theta_{g/e}$ where we defined angles $\theta_{g/e}=\pi/2 \mp \tilde{\theta}$ with $\sin\tilde{\theta}=1/\sqrt{1+D_r^2/\chi^{2}}$~\footnote{The stability of the groundstate dictates $1>2\left|D_{s}/\omega\right|+\sqrt{\left(\chi/\omega\right)^{2}+\left(D_{r}/\omega\right)^{2}}$.}. Consequently, the pseudofield is modified by the qubit as $\boldsymbol{\omega}_{g/e} = \boldsymbol{\omega}_0 \pm 2\chi \hat{e}_z$~\cite{SM}, which is illustrated in the schematics of Fig.~\ref{fig:Fig4}(a). We find that pseudospin squeezing of states $\ket{m,n}_g$ ($\ket{m,n}_e$) can be defined for the rotated pseudospin components $\hat{J}_i^{g/e}=R_y\!(\theta_{g/e})\hat{L}_iR_y^\dagger\!(\theta_{g/e})$. The corresponding squeeze factor $\xi^g_{m0} = \xi_{m0}^e \equiv \xi^\chi_{m0}$ can be obtained from Eq.~\eqref{eq:xi} upon substitution $D_r \rightarrow \sqrt{D_r^2 + \chi^2}$. For small $\chi/D_r \ll 1$, we find $\xi^\chi_{m0} \approx \xi_{m0} + c_2\chi^2$~\footnote{Factor $c_2$ is given by  $c_2 = -\frac{Ds/Dr}{8}\left(\frac{2(m+\frac{1}{2})a_+\sinh(2r_+-2r_-)}{(m+\frac{1}{2})\cosh(2r_+-2r_-)-\frac{1}{2}} - \frac{2(m+\frac{1}{2})a_-\sinh(2r_++2r_-)}{(m+\frac{1}{2})\cosh(2r_++2r_-)-\frac{1}{2}}\right)$ with $a_\pm = \frac{1}{(\omega - D_r)^2 - 4D_s^2} \pm \frac{1}{(\omega + D_r)^2 - 4D_s^2}.$}~\cite{SM}. Thus, the magnon pseudospin state, eigenmodes, and fluctuations can be modified in a controlled way via the coupling strength $\chi$ and the state of the qubit.
	
	Finally, the ground state $\ket{0,0}_g$ can be expressed in terms of a quantum superposition of multiple excited states $\ket{m,n}_e$. This compositions, for small $\chi/D_r \ll 1$, is obtained as~\cite{SM} 
	\begin{align}
		\ket{0,0}_{g}&\approx\ket{0,0}_{e} - \chi \frac{\sinh(r_+-r_-)}{2D_r}\ket{1,1}_{e}.
	\end{align}  
	Because of the nonvanishing overlap $\mathop{\vphantom{\ket{0}}}_e\!\braket{1,1}{0,0}_g\neq 0$, the qubit can be excited into the state $\ket{1,1}_e$ when it is driven at frequency $\omega_{11}$ -- the energy difference between $\ket{0,0}_g$ and $\ket{1,1}_e$~\cite{SM} -- with a probability of $c=\bigl|\mathop{\vphantom{\ket{0}}}_e\!\braket{1,1}{0,0}_g\bigr|^2$. Driving the qubit over a range of frequencies and measuring its steady state excitation reveals a nontrivial peak around $\omega_{11}$ with a height proportional to the excitation probability $c$, which we label as the contrast. This nontrivial signature stems from pseudospin squeezing and can be used to resolve the ground state quantum superposition via the described qubit spectroscopy~\cite{romling2023}. In Fig.~\ref{fig:Fig4}(b) we plot contrast $c$ as a function of $\chi$ in the small $\chi$ limit.
	
	\textit{Discussion and conclusions.---}
	Antiferromagnets as a platform hosting interacting bosonic modes offer a convenient tunability of the system parameters $\omega$, $r$, $D_r$ and $D_s$ via external magnetic fields and shape anisotropy~\cite{parvini2020, janus2023,jungwirth2016}. However, our results are applicable to other bosonic systems and theoretically introduce quantum engineering of bosonic pseudospin and its fluctuations. 
	
	As detailed in the SM~\cite{SM}, the easy-plane AFMs nickel oxide and hematite above Morin temperature can be described by a Hamiltonian in the form of $\hat{\mathcal{H}}_\mathrm{AFM}$ [Eq.~\eqref{eq:Hamiltonian2}] and hence realize the model presented here. However, a similar effective Hamiltonian can be achieved for easy-axis AFMs in the spin-flop and the canted phase~\cite{rezende2019}. While the intrasublattice interactions responsible for the magnon pseudospin squeezing are typically weaker than exchange and are often neglected, the coupling strengths $D_r$ and $D_s$ are greatly enhanced by intrinsic squeezing $\propto \sinh(2r)$ and $\propto \cosh(2r)$ respectively~\cite{SM} making the magnon pseudospin squeezing a large effect. In the case of nickel oxide, the coupling strengths are estimated to be $D_r/\omega\approx 0.48$ and $D_s/\omega\approx 0.24$. Thus, the relatively large values of $D_r$ and $D_s$ evaluated here highlight the importance of taking SNC interactions into account~\cite{SM}. Furthermore, the model we study here can be used to describe recently realized qubit-magnon platforms, such as the antiferromagnetic van der Waals magnet CrSBr with a defect acting as a spin qubit~\cite{klein2024,melendez2025}.
	
	Considering atomic ensembles as an example, spin squeezing has been mathematically proposed as an entanglement test for bosons~\cite{dalton2017}. Here, we present a concrete platform exhibiting the spin squeezing~\cite{dalton2017} along with a method of probing squeezing via qubit spectroscopy. Moreover, this platform provides squeezed Fock states~\cite{nieto1997,kral1990} which have recently been proposed as a resource for quantum error correction~\cite{korolev2024}. Because of this and the possibility of state control via a qubit, our analysis highlights new opportunities for quantum technologies. At the same time, the spin-non conserving interactions and the concomitant magnon pseudospin squeezing introduce a knob to tune and engineer quantum fluctuations, coupling to electrons and mode selection expected to be useful in designing spintronic devices~\cite{jungwirth2016,jungwirth2014,wu2013,jungwirth2011}.  
	
	\begin{acknowledgments}
		This work was funded by the Spanish Ministry of Science, Innovation and Universities-Agencia Estatal de Investigac\'{\i}on through the Grants PID2021-125894NB-I00, and CEX2023-001316-M (through the ``Mar\'{\i}a de Maeztu'' program for Units of Excellence in R\&D). A. E. R. acknowledges that the project that gave rise to these results received the support of a fellowship from ``la Caixa'' Foundation (ID 100010434) with the fellowship code LCF/BQ/DI22/11940029. A.K. was supported by the Spanish Ministry for Science and Innovation– AEI grant RYC2021 031063-I, and the German Research Foundation (DFG) via Spin+X TRR 173-268565370, project A13. F.J.G.-V also thanks the support from the ‘(MAD2D-CM)-UAM7’ project funded by Comunidad de Madrid, by the Recovery, Transformation and Resilience Plan, and by NextGenerationEU from the European Union.
	\end{acknowledgments}

	\bibliography{pseudospin_sq_lib}

\widetext
\clearpage
\setcounter{equation}{0}
\setcounter{figure}{0}
\setcounter{table}{0}
\makeatletter
\renewcommand{\theequation}{S\arabic{equation}}
\renewcommand{\thefigure}{S\arabic{figure}}

\begin{center}

	\textbf{\large Supplementary material with the manuscript Squeezing and quantum control of antiferromagnetic magnon pseudospin by} \\
	\vspace{0.3cm}
	Anna-Luisa E. R\"omling, Johannes Feist, Francisco J. García-Vidal, and Akashdeep Kamra
	\vspace{0.2cm}
\end{center}
	
	\setcounter{page}{1}
	
	\section{Derivation and diagonalization of the Hamiltonian}
	In this section, we want to derive the Hamiltonian used in the main text $\hat{\mathcal{H}}_\mathrm{AFM}$ [Eq.~(4)] from a spin model and then proceed to diagonalize it. In Sec.~\ref{sec:Hamiltonian}, we perform the derivation of the system Hamtilonian $\hat{\mathcal{H}}_\mathrm{AFM}$ [Eq.~(4)] for the easy-plane antiferromagnet (AFM) Nickel Oxide and discuss the origin of coupling strengths $D_r$ and $D_s$. In Sec.~\ref{sec:diagonalization}, we proceed by diagonalizing $\hat{\mathcal{H}}_\mathrm{AFM}$ making use of the pseudospin formalism and discussing the transformations and operators involved in the diagonalization. 
	\subsection{Spin Hamiltonian of easy-plane antiferromagnet \label{sec:Hamiltonian}}
	Here, we derive the Hamiltonian $\hat{\mathcal{H}}_\mathrm{AFM}$ [Eq.~(4) in the main text] from a spin model for the easy-plane AFM Nickel Oxide (NiO). We start from classical ground state of an AFM, called N\'eel state, that consists of two fully and oppositely polarized sublattices \al{below the Néel temperature. NiO has a hard-axis and easy-axis anisotropy. The hard-axis anisotropy is along the $\langle111\rangle$ axis and forces the spins into the $\{111\}$ planes~\cite{rezende2019, hutchings1972}. The easy-axis anisotropy is along the $\langle11\bar{2}\rangle$ axes. There are three equivalent $\langle11\bar{2}\rangle$ directions in the $\{111\}$ planes~\cite{rezende2019}.} Here, we denote the spin-down (up) sublattice as $A$ ($B$).  Following~\cite{rezende2019, rezende2016,hutchings1972}, we start with the Hamiltonian describing a bipartite lattice AFM with easy-plane anisotropy and applied magnetic field ($\hbar=1$)
	\begin{align}
		\mathcal{H}_\mathrm{NiO}  = &|\gamma| H_{0}\sum_{i\in A,j\in B}\left(\hat{S}_{i}^{z^{\prime}}+\hat{S}_{j}^{z^{\prime}}\right)+2J\sum_{\left\langle i,j\right\rangle }\hat{\boldsymbol{S}}_{i}\cdot\hat{\boldsymbol{S}}_{j}+K_{x}\sum_{i\in A,j\in B}\left[\left(\hat{S}_{i}^{x}\right)^{2}+\left(\hat{S}_{j}^{x}\right)^{2}\right]\nonumber \\ 
		&-K_{z}\sum_{i\in A,j\in B}\left[\left(\hat{S}_{i}^{z^{\prime}}\right)^{2}+\left(\hat{S}_{j}^{z^{\prime}}\right)^{2}\right], \label{eq:Hamiltonian_AFM_spin}
	\end{align}
	with $\hat{S}_{i/j}^{\pm} = \hat{S}_{i/j}^x \pm i \hat{S}_{i/j}^y$, the gyromagnetic ratio $\gamma<0$, $J$ denoting the exchange interaction of nearest neighbors belonging to different sublattices, the easy-plane anisotropy $K_x$ and easy-axis anisotropy $K_z$. \al{Here, $z^\prime$ denotes one of the easy directions along the $\langle11\bar{2}\rangle$ axes. We also consider an applied magnetic field along one of the $z^\prime$-directions, denoted by $\boldsymbol{H} = H_0\hat{\boldsymbol{e}}_{z^\prime}$.} Note here that the index $i$ ($j$) marks lattice sites belonging to sublattice $A$ ($B$).
	
	In order to quantize our Hamiltonian $\hat{\mathcal{H}}_\mathrm{NiO}$ [Eq.~\eqref{eq:Hamiltonian_AFM_spin}], we use the following linearized Holstein-Primakoff (HP) transformations for lattice sites $i\in A$ and $j \in B$~\cite{holstein1940, rezende2019}
	\begin{align}
		\begin{split}
			\hat{S}_{i}^{+} & =\sqrt{\frac{2S}{N}}\sum_{\boldsymbol{k}}e^{-i\boldsymbol{k}\cdot\boldsymbol{r}_i}\hat{a}_{\boldsymbol{k}}^\dagger,\\
			\hat{S}_{i}^{-} & =\sqrt{\frac{2S}{N}}\sum_{\boldsymbol{k}}e^{i\boldsymbol{k}\cdot\boldsymbol{r}_i}\hat{a}_{\boldsymbol{k}},\\
			\hat{S}_{i}^{z^{\prime}} & =-S+\frac{1}{N}\sum_{\boldsymbol{k},\boldsymbol{k}^{\prime}}e^{-i\left(\boldsymbol{k}-\boldsymbol{k}^{\prime}\right)\cdot\boldsymbol{r}_i}\hat{a}_{\boldsymbol{k}}^{\dagger}\hat{a}_{\boldsymbol{k}^{\prime}},\\
		\end{split}
		\begin{split}
			\hat{S}_{j}^{+} & =\sqrt{\frac{2S}{N}}\sum_{\boldsymbol{k}}e^{i\boldsymbol{k}\cdot\boldsymbol{r}_j}\hat{b}_{\boldsymbol{k}},\\
			\hat{S}_{j}^{-} & =\sqrt{\frac{2S}{N}}\sum_{\boldsymbol{k}}e^{-i\boldsymbol{k}\cdot\boldsymbol{r}_j}\hat{b}_{\boldsymbol{k}}^\dagger,\\
			\hat{S}_{j}^{z\prime} & =S-\frac{1}{N}\sum_{\boldsymbol{k},\boldsymbol{k}^{\prime}}e^{-i\left(\boldsymbol{k}-\boldsymbol{k}^{\prime}\right)\cdot\boldsymbol{r}_j}\hat{b}_{\boldsymbol{k}}^{\dagger}\hat{b}_{\boldsymbol{k}^{\prime}},\label{eq:HP}
		\end{split} 
	\end{align}
	where $S$ denotes the individual total spin of each lattice site and $N$ the number of sublattice spins. The quantization axes are different for the two sublattice because of their opposite spin orientation. The operators $\hat{a}_{\boldsymbol{k}}$ ($\hat{b}_{\boldsymbol{k}}$) represent the annihilation operator of a spin-up (down) excitation on sublattice $A$ ($B$). They fulfil the bosonic commutation relations $\left[\hat{a}_{\boldsymbol{k}}, \hat{a}_{\boldsymbol{k^\prime}}^\dagger\right]=\delta_{\boldsymbol{k}\boldsymbol{k}^\prime}$, $\left[\hat{b}_{\boldsymbol{k}}, \hat{b}_{\boldsymbol{k^\prime}}^\dagger\right]=\delta_{\boldsymbol{k}\boldsymbol{k}^\prime}$ and $\left[\hat{a}_{\boldsymbol{k}}, \hat{b}_{\boldsymbol{k^\prime}}^\dagger\right]=0$ and will be denoted spin-up (down) sublattice magnons $\hat{a}_{\boldsymbol{k}}$ ($\hat{b}_{\boldsymbol{k}}$). Note that the linearized HP transformations are only valid if the average number of excitations is small in comparison with the spin at each lattice site $S$. Using the linear HP transformations [Eq.~\eqref{eq:HP}], we arrive at the Hamiltonian
	\begin{align}
		\hat{\mathcal{H}}_\mathrm{NiO} & =\sum_{\boldsymbol{k}}\frac{A}{2}\hat{a}_{\boldsymbol{k}}^{\dagger}\hat{a}_{\boldsymbol{k}}+\frac{B}{2}\hat{b}_{\boldsymbol{k}}^{\dagger}\hat{b}_{\boldsymbol{k}}+C_{\boldsymbol{k}}\hat{a}_{\boldsymbol{k}}\hat{b}_{-\boldsymbol{k}}+D\left(\hat{a}_{\boldsymbol{k}}\hat{a}_{-\boldsymbol{k}}+\hat{b}_{\boldsymbol{k}}\hat{b}_{-\boldsymbol{k}}\right) + \mathrm{h.c.}, \label{eq:Hamiltonian_boson}
	\end{align}
	with 
	\begin{align}
		\begin{split}
			A & =|\gamma|\left(H_{E}+\frac{H_{Ax}}{2}+H_{Az}+H_{0}\right),\\
			B & =|\gamma|\left(H_{E}+\frac{H_{Ax}}{2}+H_{Az}-H_{0}\right),\\
		\end{split}
		\begin{split}
			C_{\boldsymbol{k}} & =|\gamma|\gamma_{\boldsymbol{k}}H_{E},\\
			D & =|\gamma|\frac{H_{Ax}}{4}, \label{eq:ABCD}
		\end{split}
	\end{align}
	and $H_{E} =2SzJ/|\gamma|$, $H_{Ax}=2SK_{x}/|\gamma|$, $H_{Az}=2SK_{z}/|\gamma|$, and $\gamma_{\boldsymbol{k}} =\cos\left(a\left|\boldsymbol{k}\right|/2\right)$. Here, $z$ is the number of nearest neighbors and $a$ the lattice constant of NiO. 
	
	Considering a nanomagnet where the $\boldsymbol{k}\neq\boldsymbol{0}$ modes are well separated from the Kittel mode $\boldsymbol{k}=\boldsymbol{0}$~\cite{skogvoll2021}, we focus on the homogeneous mode $\boldsymbol{k}=\boldsymbol{0}$ here. We can therefore drop the index $\boldsymbol{k}$ in the operators and coefficients and define $\hat{a}\equiv\hat{a}_{\boldsymbol{0}}$, $\hat{b}\equiv\hat{b}_{\boldsymbol{0}}$
	and $C\equiv C_{\boldsymbol{0}}$ where $\gamma_{\boldsymbol{0}}=1$. Additionally assuming $H_0=0$, such that $A=B$ [Eq.~\eqref{eq:ABCD}], our Hamiltonian $\mathcal{H}_\mathrm{NiO}$ [Eq.~\eqref{eq:Hamiltonian_boson}] becomes
	\begin{align}
		\hat{\mathcal{H}}_\mathrm{NiO} & =A\hat{a}^{\dagger}\hat{a}+B\hat{b}^{\dagger}\hat{b}+C\left(\hat{a}\hat{b}+\hat{a}^{\dagger}\hat{b}^{\dagger}\right)+D\left(\hat{a}^{2}+\hat{b}^{2}+\hat{a}^{\dagger2}+\hat{b}^{\dagger2}\right).\label{eq:H_AFM}
	\end{align}
	where, for now, we assume the parameters $A$, $C$ and $D$ to be real. We define spin-up and spin-down magnon operators $\hat{\alpha}$ and $\hat{\beta}$ via the two-mode Bogoliubov transformation~\cite{kamra2019}
	\begin{align}
		\left(\begin{array}{c}
			\hat{\alpha}\\
			\hat{\beta}^{\dagger}
		\end{array}\right) & =\left(\begin{array}{cc}
			u & v\\
			v & u
		\end{array}\right)\left(\begin{array}{c}
			\hat{a}\\
			\hat{b}^{\dagger}
		\end{array}\right),\label{eq:Bogoliubov}
	\end{align}
	with the coefficients 
	\begin{align}
		\begin{split}
			u & =\frac{1}{\sqrt{2}}\sqrt{\frac{A}{\omega}+1},
		\end{split}
		\begin{split}
			v & =\frac{1}{\sqrt{2}}\sqrt{\frac{A}{\omega}-1},\label{eq:u}
		\end{split}
	\end{align}
	and the energy $\omega =\sqrt{A^{2}-C^{2}}$. The Bogoliubov coefficients $u$ and $v$ [Eq.~\eqref{eq:u}] define intrinsic two-mode squeezing $r$ by $\tanh(r)=v/u$. This squeezing stems from the strong inter-sublattice exchange interaction and is typically large in AFMs. Magnons, represented by $\hat{\alpha}$ and $\hat{\beta}$, are therefore intrinsically squeezed. In the new basis [Eq.~\eqref{eq:Bogoliubov}], the Hamiltonian $\hat{\mathcal{H}}_\mathrm{NiO}$ [Eq.~\eqref{eq:H_AFM}] reads
	\begin{align}
		\hat{\mathcal{H}}_\mathrm{AFM} & =\omega\left(\hat{\alpha}^{\dagger}\hat{\alpha}+\hat{\beta}^{\dagger}\hat{\beta}\right)-D_{r}\left(\hat{\alpha}\hat{\beta}^{\dagger}+\hat{\alpha}^{\dagger}\hat{\beta}\right)+D_{s}\left(\hat{\alpha}^{2}+\hat{\alpha}^{\dagger2}+\hat{\beta}^{2}+\hat{\beta}^{\dagger2}\right)+\left(\omega-A\right),\label{eq:H_AFM_sq}
	\end{align}
	where we defined $D_{r}=2D\sinh(2r)$ and $D_{s}=D\cosh(2r)$. The form of the antiferromagnetic Hamiltonian [Eq.~\eqref{eq:H_AFM_sq}] is the form we used in the main text which is why we denote the Hamiltonian by $\hat{\mathcal{H}}_\mathrm{AFM}$ from now on. We show in the following section that the parameter $D_r$ causes a hybridization of modes $\hat{\alpha}$ and $\hat{\beta}$ and hence a \textit{rotation} of pseudospin on the Bloch sphere whereas $D_s$ induces \textit{squeezing} of the hybridized modes and hence the pseudospin components.
	Finally, using the following effective values $H_{E} =9684\,\text{kOe}$, $H_{Ax}  =6.35\,\text{kOe}$, $H_{Az} =0.11\,\text{kOe}$~\cite{rezende2019} and $H_{0} =0$, we find the coefficients $A = B =|\gamma|\cdot9687.285\,\text{kOe}$, $C =|\gamma|\cdot9684\,\text{kOe}$ and $D =|\gamma|\cdot1.5875\,\text{kOe}$. Therefore, $D$ is four orders of magnitude smaller than typical values of $A$, $B$, and $C$, thereby motivating why it has often been disregarded in previous studies. However, $D_r$ and $D_s$ are enhanced by intrinsic squeezing $r$. We estimate $\omega\approx |\gamma|\cdot252.26\,\text{kOe} = 2\pi\cdot 0.7\,\text{THz}$, $r\approx 2.17$, $D_r/\omega \approx 0.48$ and  $D_s/\omega \approx 0.24$. 
	\subsection{Eigenstates of the bare antiferromagnet \label{sec:diagonalization}}
	In this section, we diagonalize the Hamiltonian $\hat{\mathcal{H}}_\mathrm{AFM}$ [Eq.~\eqref{eq:H_AFM_sq}] via the pseudospin formalism~\cite{kamra2020a, dalton2017} and a one-mode squeeze transformation. We start from a reduced Hamiltonian, setting $D_s=0$ in $\hat{\mathcal{H}}_\mathrm{AFM}$ [Eq.~\eqref{eq:H_AFM_sq}], which results in 
	\begin{equation}
		\hat{\mathcal{H}}_\mathrm{hyb} = \omega\left(\hat{\alpha}^\dagger\hat{\alpha} + \hat{\beta}^\dagger\hat{\beta}\right) - D_r\left(\hat{\alpha}\hat{\beta}^\dagger + \hat{\alpha}^\dagger \hat{\beta}\right).\label{eq:H_hyb}
	\end{equation}
	The coherent coupling between the spin-up and spin-down magnon $\hat{\alpha}$ and  $\hat{\beta}$ proportional to $\propto D_r$ leads to a hybridization of magnons. 
	
	As the next step, we introduce the symmetric and antisymmetric modes $\hat{\psi}_s$ and $\hat{\psi}_a$ as
	\begin{align}
		\begin{split}
			\hat{\psi}_{s} & =\frac{1}{\sqrt{2}}\left(\hat{\alpha}+\hat{\beta}\right),
		\end{split}
		\begin{split}
			\hat{\psi}_{a} & =\frac{1}{\sqrt{2}}\left(-\hat{\alpha}+\hat{\beta}\right),\label{eq:psi_sa}
		\end{split}
	\end{align}
	which diagonalize $\hat{\mathcal{H}}_{\mathrm{hyb}}$ [Eq.~\eqref{eq:H_hyb}] as 
	\begin{align}
		\hat{\mathcal{H}} & =\omega_{s}\psi_{s}^{\dagger}\psi_{s}+\omega_{a}\psi_{a}^{\dagger}\psi_{a}+\left(\omega-A\right),\label{eq:H_hyb_diag}
	\end{align}
	with the energies
	\begin{align}
		\begin{split}
			\omega_{s} & =\omega-D_{r},
		\end{split}
		\begin{split}
			\omega_{a} & =\omega+D_{r}.
		\end{split}
	\end{align}
	The symmetric and antisymmetric modes $\hat{\psi}_s$ and $\hat{\psi}_a$ [Eq.~\eqref{eq:psi_sa}] correspond to maximal hybridization of the magnons $\hat{\alpha}$ and $\hat{\beta}$. This is because magnons $\hat{\alpha}$ and $\hat{\beta}$ are degenerate here. The eigenmodes of $\hat{\mathcal{H}}_{\mathrm{hyb}}$ can also be treated within a pseudospin framework defining the pseudospin components~\cite{kamra2020a} 
	\begin{align}
		\begin{split}
			\hat{L}_{0} & =\frac{1}{2}\left(\hat{\alpha}^{\dagger}\hat{\alpha}+\hat{\beta}^{\dagger}\hat{\beta}\right),
		\end{split}
		\begin{split}
			\hat{\boldsymbol{L}} & =\frac{1}{2}\left(\begin{array}{c}
				\hat{\alpha}\hat{\beta}^{\dagger}+\hat{\alpha}^{\dagger}\hat{\beta}\\
				i\hat{\alpha}\hat{\beta}^{\dagger}-i\hat{\alpha}^{\dagger}\hat{\beta}\\
				\hat{\alpha}^{\dagger}\hat{\alpha}-\hat{\beta}^{\dagger}\hat{\beta}
			\end{array}\right). \label{eq:pseudospin_sm}
		\end{split}
	\end{align}
	The Hamiltonian $\hat{\mathcal{H}}_\mathrm{hyb}$ can be reformulated using pseudospin operators $\hat{L}_i$, $i\in\{0,1,2,3\}$ as
	\begin{align}
		\hat{\mathcal{H}}_{\text{hyb}}& =2\omega\hat{L}_{0}-\boldsymbol{\omega}^{0}\cdot\hat{\boldsymbol{L}}+\left(\omega-A\right),\label{eq:H_hyb_ps}
	\end{align}
	with pseudofield $\boldsymbol{\omega}^0 = 2D_r\hat{\boldsymbol{e}}_x$. In the peudospin framework, we can picture the eigenmodes on a Bloch unit sphere, where the poles correspond to the $\hat{\alpha}$ and $\hat{\beta}$ magnons. The hybridized modes are then characterized by the pseudofield $\boldsymbol{\omega}^0$ and correspond to the intersection of the pseudosfield with the Bloch sphere. The eigenmodes $\hat{\psi}_s$ and $\hat{\psi}_a$ can be obtained from the following transformation~\cite{kamra2020a} 
	\begin{align}
		\left(\begin{array}{c}
			\hat{\psi}_{s}\\
			\hat{\psi}_{a}
		\end{array}\right) & =\left(\begin{array}{cc}
			\cos(\frac{\theta}{2}) & \sin(\frac{\theta}{2})\\
			-\sin(\frac{\theta}{2}) & \cos(\frac{\theta}{2})
		\end{array}\right)\left(\begin{array}{c}
			\hat{\alpha}\\
			\hat{\beta}
		\end{array}\right),\label{eq:pseudospin_sa}
	\end{align}
	with the rotation angle $\cos\theta=\omega^0_z/\left|\boldsymbol{\omega}^0\right|$. Here, this angle is $\theta = \pi/2$. Finally, the transformation Eq.~\eqref{eq:pseudospin_sa} can be expressed via a rotation operator
	\begin{align}
		\hat{R}_{y}\!\left(\theta\right) & =\exp(-i\theta\hat{L}_{y}),\label{eq:R_y_theta}
	\end{align}
	such that the hybridized modes read 
	\begin{align}
		\begin{split}
			\hat{\psi}_s & =\hat{R}_{y}\!\left(\frac{\pi}{2}\right)\hat{\alpha}\hat{R}_{y}^{\dagger}\!\left(\frac{\pi}{2}\right),\\
		\end{split}
		\begin{split}
			\hat{\psi}_a & =\hat{R}_{y}\!\left(\frac{\pi}{2}\right)\hat{\beta}\hat{R}_{y}^{\dagger}\!\left(\frac{\pi}{2}\right), \label{eq:rotation_sa}
		\end{split}
	\end{align}
	\al{which have bosonic commutation relations $[\hat{\psi}_{s/a},\hat{\psi}^\dagger_{s/a}]=1$ and $[\hat{\psi}_s,\hat{\psi}_a]=0$. The relations in Eq.~\eqref{eq:rotation_sa} can be proven straightforwardly with the Baker-Campbell-Hausdorff relation for matrix exponential and mathematical induction. The proof is performed in detail in Appendix \ref{sec:Rotation-transformation}.} The rotation operator $\hat{R}_y\!\left(\theta\right)$ [Eq.~\eqref{eq:R_y_theta}] will be convenient when expressing the eigenmodes of $\hat{\mathcal{H}}_\mathrm{AFM}$ [Eq.~\eqref{eq:H_AFM_sq}] and especially important when addressing state control via a qubit in Sec.~\ref{sec:qubit}. 
	
	Now using the hybridized modes $\hat{\psi}_s$ and $\hat{\psi}_a$ [Eq.~\eqref{eq:psi_sa}], we can transform the full Hamiltonian $\hat{\mathcal{H}}_\mathrm{AFM}$ [Eq.~\eqref{eq:H_AFM_sq}] and obtain the following expression 
	\begin{align}
		\hat{\mathcal{H}}_\mathrm{AFM} & =\omega_{s}\hat{\psi}_{s}^{\dagger}\hat{\psi}_{s}+\omega_{a}\hat{\psi}_{a}^{\dagger}\hat{\psi}_{a}+D_{s}\left(\hat{\psi}_{s}^{2}+\hat{\psi}_{s}^{\dagger2}+\hat{\psi}_{a}^{2}+\hat{\psi}_{a}^{\dagger2}\right)+\left(\omega-A\right).\label{eq:H_quad_sa}
	\end{align}
	In the current form of $\hat{\mathcal{H}}_\mathrm{AFM}$ [Eq.~\eqref{eq:H_quad_sa}], the hybridized modes $\hat{\psi}_{s/a}$ are decoupled and we can diagonalize their respective contribution to the Hamiltonian separately. We can see that because of the squared terms $\propto D_s$, the Hamiltonian describes two non-interacting one-mode squeezed harmonic oscillators. Therefore the final step is to
	perform a one-mode squeezing transformation for both modes, which results in the following diagonalized version of the Hamiltonian
	\begin{align}
		\hat{\mathcal{H}}_\mathrm{AFM} & =\omega_{+}\hat{\psi}_{+}^{\dagger}\hat{\psi}_{+}+\omega_{-}\hat{\psi}_{-}^{\dagger}\hat{\psi}_{-}+\frac{\omega_{+}+\omega_{-}-2A}{2},\label{eq:H_diag}
	\end{align}
	with the one-mode squeezed hybridized modes
	\begin{align}
		\begin{split}
			\hat{\psi}_{+} & =\cosh(r_{+})\hat{\psi}_{s}+\sinh(r_{+})\hat{\psi}_{s}^{\dagger},\\
		\end{split}
		\begin{split}
			\hat{\psi}_{-} & =\cosh(r_{-})\hat{\psi}_{a}+\sinh(r_{-})\hat{\psi}_{a}^{\dagger},\label{eq:psi_sq}
		\end{split}
	\end{align}
	the one mode-squeezing factors 
	\begin{align}
		\begin{split}
			\tanh(2r_{+}) & =\frac{2\left|D_{s}\right|}{\omega-D_{r}},
		\end{split}
		\begin{split}
			\tanh(2r_{-}) & =\frac{2\left|D_{s}\right|}{\omega+D_{r}},\label{eq:rs_ra}
		\end{split}
	\end{align}
	and the eigenenergies 
	\begin{align}
		\begin{split}
			\omega_{+} & =\sqrt{\omega^{2}-4CD-4D^{2}},\\
		\end{split}
		\begin{split}
			\omega_{-} & =\sqrt{\omega^{2}+4CD-4D^{2}}.
		\end{split}
	\end{align}
	Here, we want to mention that the one-mode squeeze transformation [Eq.~\eqref{eq:psi_sq}] can be conveniently expressed via the one-mode squeeze operators
	\begin{equation}
		\hat{S}_{\pm}\!\left(r_{\pm}\right)=\exp( \frac{r_{\pm}}{2}\left[\left(\hat{\psi}_{s/a}\right)^{2}-\left(\hat{\psi}_{s/a}\right)^{2}\right]),\label{eq:squeeze-sa}
	\end{equation}
	such that the eigenmodes read $\hat{\psi}_{+} = \hat{S}_{+}\!\left(r_{+}\right) \hat{\psi}_{s}\hat{S}_{+}^{\dagger}\!\left(r_{+}\right)$ and $\hat{\psi}_{-} = \hat{S}_{-}\!\left(r_{-}\right) \hat{\psi}_{a}\hat{S}_{-}^{\dagger}\!\left(r_{-}\right)$~\cite{breuer2007}. \al{With the relations from Eq.~\eqref{eq:rotation_sa}, the eigenmodes can be written as 
		\begin{align}
			\hat{\psi}_{+} = \hat{S}_{+}\!\left(r_{+}\right) \hat{R}_{y}\!\left(\frac{\pi}{2}\right)\hat{\alpha}\hat{R}_{y}^{\dagger}\!\left(\frac{\pi}{2}\right)\hat{S}_{+}^{\dagger}\!\left(r_{+}\right),\label{eq:psi_+}\\
			\hat{\psi}_{-} = \hat{S}_{-}\!\left(r_{-}\right) \hat{R}_{y}\!\left(\frac{\pi}{2}\right)\hat{\beta}\hat{R}_{y}^{\dagger}\!\left(\frac{\pi}{2}\right)\hat{S}_{-}^{\dagger}\!\left(r_{-}\right).\label{eq:psi_-}
	\end{align}}
	Finally, a general eigenstate of the Hamiltonian $\hat{\mathcal{H}}_\mathrm{AFM}$ [Eq.~\eqref{eq:H_AFM_sq}] with $m$ ($n$) excitations in the $\hat{\psi}_+$ ($\hat{\psi}_-$) mode can be expressed as 
	\begin{align}
		\ket{m_+,n_-} & =\hat{S}_{+}\!\left(r_{+}\right)\hat{S}_{-}\!\left(r_{-}\right)\hat{R}_y\!\left(\frac{\pi}{2}\right)\ket{m_\alpha,n_\beta},\label{eq:Fock_without_qubit}
	\end{align}
	where $\ket{m_\alpha,n_\beta}$ denotes the eigenstates of $\omega\left(\hat{\alpha}^{\dagger}\hat{\alpha}+\hat{\beta}^{\dagger}\hat{\beta}\right)$ with $m$ ($n$) denoting the number spin-up (down) magnonic excitations. \al{The relation in Eq.~\eqref{eq:Fock_without_qubit} follows from first considering the vacuum $\ket{0_+,0_-}$ which has to fulfil $\hat{\psi}_{\pm}\ket{0_+,0_-} = 0$. This can be shown straightforwardly 
		\begin{align}
			\hat{\psi}_{+}\ket{0_+,0_-} &= \hat{S}_{+}\!\left(r_{+}\right) \hat{R}_{y}\!\left(\frac{\pi}{2}\right)\hat{\alpha}\hat{R}_{y}^{\dagger}\!\left(\frac{\pi}{2}\right)\hat{S}_{+}^{\dagger}\!\left(r_{+}\right)\hat{S}_{+}\!\left(r_{+}\right)\hat{S}_{-}\!\left(r_{-}\right)\hat{R}_y\!\left(\frac{\pi}{2}\right)\ket{0_\alpha,0_\beta}\\
			& = \hat{S}_{+}\!\left(r_{+}\right) \hat{R}_{y}\!\left(\frac{\pi}{2}\right)\hat{\alpha}\hat{R}_{y}^{\dagger}\!\left(\frac{\pi}{2}\right)\hat{S}_{-}\!\left(r_{-}\right)\hat{R}_y\!\left(\frac{\pi}{2}\right)\ket{0_\alpha,0_\beta}\\
			&= \hat{S}_{+}\!\left(r_{+}\right) \hat{S}_{-}\!\left(r_{-}\right)\hat{R}_{y}\!\left(\frac{\pi}{2}\right)\hat{\alpha}\hat{R}_{y}^{\dagger}\!\left(\frac{\pi}{2}\right)\hat{R}_y\!\left(\frac{\pi}{2}\right)\ket{0_\alpha,0_\beta}\\
			&= \hat{S}_{+}\!\left(r_{+}\right) \hat{S}_{-}\!\left(r_{-}\right)\hat{R}_{y}\!\left(\frac{\pi}{2}\right)\hat{\alpha}\ket{0_\alpha,0_\beta}\\
			&=0,
		\end{align}
		where we used the property of unitary operators $\hat{S}_{+}^{\dagger}\!\left(r_{+}\right)\hat{S}_{+}\!\left(r_{+}\right)=1$ and $\hat{R}_{y}^{\dagger}\!\left(\frac{\pi}{2}\right)\hat{R}_y\!\left(\frac{\pi}{2}\right)=1$, the commutator $[\hat{R}_{y}\!\left(\frac{\pi}{2}\right)\hat{\alpha}\hat{R}_{y}^{\dagger}\!\left(\frac{\pi}{2}\right),\hat{S}_{-}\!\left(r_{-}\right)] = [\hat{\psi}_s,\hat{S}_{-}\!\left(r_{-}\right)]=0$, where we used the relations from Eqs.~\eqref{eq:rotation_sa} and \eqref{eq:squeeze-sa} and $\hat{\alpha}\ket{0_\alpha,0_\beta}=0$. From this then follows that 
		\begin{align}
			\ket{m_+,n_-} &= \frac{1}{\sqrt{m!n!}}\hat{\psi}_+^{\dagger m}\hat{\psi}_+^{\dagger m} \ket{0_+,0_-}\\ &=\frac{1}{\sqrt{m!n!}}\hat{S}_{+}\!\left(r_{+}\right) \hat{R}_{y}\!\left(\frac{\pi}{2}\right)\hat{\alpha}^{\dagger m}\hat{R}_{y}^{\dagger}\!\left(\frac{\pi}{2}\right)\hat{S}_{+}^{\dagger}\!\left(r_{+}\right) \times \nonumber\\
			&~~~~\times \hat{S}_{-}\!\left(r_{-}\right) \hat{R}_{y}\!\left(\frac{\pi}{2}\right)\hat{\beta}^{\dagger m}\hat{R}_{y}^{\dagger}\!\left(\frac{\pi}{2}\right)\hat{S}_{-}^{\dagger}\!\left(r_{-}\right)\hat{S}_{+}\!\left(r_{+}\right)\hat{S}_{-}\!\left(r_{-}\right)\hat{R}_y\!\left(\frac{\pi}{2}\right)\ket{0_\alpha,0_\beta}\\
			& =\frac{1}{\sqrt{m!n!}}\hat{S}_{+}\!\left(r_{+}\right)\hat{S}_{-}\!\left(r_{-}\right)\hat{R}_y\!\left(\frac{\pi}{2}\right)\hat{\alpha}^{\dagger m}\hat{\beta}^{\dagger n}\ket{0_\alpha,0_\beta}\\
			&=\hat{S}_{+}\!\left(r_{+}\right)\hat{S}_{-}\!\left(r_{-}\right)\hat{R}_y\!\left(\frac{\pi}{2}\right)\ket{m_\alpha,n_\beta},
		\end{align}
		which shows the relation in Eq.~\eqref{eq:Fock_without_qubit}. Note that we also used the commutators $[\hat{R}_{y}\!\left(\frac{\pi}{2}\right)\hat{\beta}\hat{R}_{y}^{\dagger}\!\left(\frac{\pi}{2}\right),\hat{S}_{+}\!\left(r_{+}\right)] = [\hat{\psi}_a,\hat{S}_{+}\!\left(r_{+}\right)]=0$ and $[\hat{S}_{+}\!\left(r_{+}\right),\hat{S}_{-}\!\left(r_{-}\right)] =0$.}
	\section{Pseudospin squeezing\label{sec:pseudospin_squeezing}}
	In Sec.~\ref{sec:diagonalization}, we introduce the concept of pseudospin $\hat{\boldsymbol{L}}$. The pseudospin components fulfil commutation relations $[\hat{L}_i,\hat{L}_j] = i\epsilon_{ijk}\hat{L}_k$ with the Levi-Civita symbol $\epsilon_{ijk}$. Because of that, pseudospin components obey Heisenberg uncertainty relation $\langle\Delta\hat{L}_y^2\rangle\langle\Delta\hat{L}_z^2\rangle \geq |\langle\hat{L}_x\rangle|^2/4$. In this section, we demonstrate that pseudospin component $\hat{L}_y$ is squeezed according to Heisenberg uncertainty in an eigenstate $\ket{m_+,n_-}$ [Eq.~\eqref{eq:Fock_without_qubit}]. 
	\subsection{Pseudospin squeezing in a general excited state}
	In this subsection, we evaluate the expectation values and variances of pseudospin components $\hat{L}_{i}$ for $i=x,y,z$ for a general
	eigenstate $\ket{m_+,n_-}$ [Eq.~\eqref{eq:Fock_without_qubit}] with $m$ ($n$) excitation in eigenmode $\hat{\psi}_+$ ($\hat{\psi}_-$) [Eq.~\eqref{eq:psi_sq}]. We denote the expectation values by $\langle \hat{L}_{i}\rangle _{mn} = \bra{m_+,n_-}\hat{L}_i\ket{m_+,n_-}$ and the variances $\langle \Delta\hat{L}_{i}^2\rangle _{mn} = \bra{m_+,n_-}\hat{L}_i^2\ket{m_+,n_-}-\langle \hat{L}_{i}\rangle _{mn}^2$. We find the following expectation values
	\begin{align}
		\left\langle \hat{L}_{x}\right\rangle _{mn} & =\frac{m\left(\cosh^2\!r_{+}+\sinh^2\!r_{+}\right)+\sinh^2\!r_{+}}{2}-\frac{n\left(\cosh^2\!r_{-}+\sinh^2\!r_{-}\right)+\sinh^2\!r_{-}}{2},\label{eq:Lx_mn}
	\end{align}
	and $\langle \hat{L}_{y}\rangle_{mn} = \langle \hat{L}_{z}\rangle_{mn} =0$. The corresponding variances of interest read 
	\begin{align}
		\left\langle \Delta\hat{L}_{y}^{2}\right\rangle _{mn} & =\frac{2mn+m+n}{4}\left(\cosh\!r_{+}\cosh\!r_{-}-\sinh\!r_{+}\sinh\!r_{-}\right)^{2}\nonumber \\
		& +\frac{2mn+m+n+1}{4}\left(\sinh\!r_{+}\cosh\!r_{-}-\sinh\!r_{-}\cosh\!r_{+}\right)^{2},\label{eq:Ly_mn}\\
		\left\langle \Delta\hat{L}_{z}^{2}\right\rangle _{mn} & =\frac{2mn+m+n}{4}\left(\cosh\!r_{+}\cosh\!r_{-}+\sinh\!r_{+}\sinh\!r_{-}\right)^{2}\nonumber \\
		& +\frac{2mn+m+n+1}{4}\left(\sinh\!r_{+}\cosh\!r_{-}+\sinh\!r_{-}\cosh\!r_{+}\right)^{2}. \label{eq:Lz_mn}
	\end{align}
	Neglecting the intrasublattice coupling $D_{s}$, we find
	\begin{align}
		\left\langle \hat{L}_{x}\right\rangle _{mn} & =\frac{m-n}{2},\\
		\left\langle \Delta\hat{L}_{y}^{2}\right\rangle _{mn} & =\frac{2mn+m+n}{4},\\
		\left\langle \Delta\hat{L}_{z}^{2}\right\rangle _{mn} & =\frac{2mn+m+n}{4},
	\end{align}
	such that $\langle \Delta\hat{L}_{y}^{2}\rangle _{mn}=\langle \Delta\hat{L}_{z}^{2}\rangle _{mn}$. From this, we can deduce that if $D_s=0$, there is no squeezing present since the variances of $\hat{L}_y$ and $\hat{L}_z$ are the same. 
	
	As in the main text, let us analyze a simpler case with $m\neq0$, $n=0$ and $D_{s}\neq0$. We find that the expectation value of $\hat{L}_x$ and the variances of $\hat{L}_y$ and $\hat{L}_z$ [Eqs.~\eqref{eq:Lx_mn}--\eqref{eq:Lz_mn}] reduce to the following expressions
	\begin{align}
		\left\langle \hat{L}_{x}\right\rangle _{m0} & =\frac{m+\frac{1}{2}}{2}\cosh2r_{+}-\frac{\cosh2r_{-}}{4},\label{eq:Lx}\\
		\left\langle \Delta\hat{L}_{y}^{2}\right\rangle _{m0} & =\frac{m+\frac{1}{2}}{4}\cosh\left(2r_{+}-2r_{-}\right)-\frac{1}{8},\label{eq:varLy}\\
		\left\langle \Delta\hat{L}_{z}^{2}\right\rangle _{m0} & =\frac{m+\frac{1}{2}}{4}\cosh\left(2r_{+}+2r_{-}\right)-\frac{1}{8}.\label{eq:varLz}
	\end{align}
	After some algebra, we find the following uncertainty relation
	\begin{align}
		\left\langle \Delta\hat{L}_{y}^{2}\right\rangle _{m0}\left\langle \Delta\hat{L}_{z}^{2}\right\rangle _{m0} & =\frac{1}{4}\left|\left\langle \hat{L}_{x}\right\rangle \right|_{m0}^{2}+\frac{m^{2}+m}{8}\left(\cosh\!4r_{-}-1\right),\label{eq:Heisenberg}
	\end{align}
	so that the state we consider here is not minimum uncertainty.
	We check if $\left\langle \Delta\hat{L}_{y}^{2}\right\rangle _{m0}<\frac{1}{2}\left|\left\langle \hat{L}_{x}\right\rangle \right|_{m0}$
	is fulfilled and find that this condition is equivalent to 
	\begin{align}
		\left(2m+1\right)\left(\cosh\!2r_{+}-\cosh(2r_{+}-2r_{-})\right)> & \cosh\!2r_{-}-1.
	\end{align}
	This is fulfilled if $r_{-}>0$ and $r_{+}>r_{-}$. This means that $L_{y}$
	is squeezed with respect to $L_{x}$ and that $\langle\Delta\hat{L}_y^2\rangle < \langle\Delta\hat{L}_z^2\rangle$. We can quantify squeezing by the ellipticity of Heisenberg's uncertainty relation [Eq.~\eqref{eq:Heisenberg}]~\cite{gerry2005}, defining
	\begin{align}
		\left\langle \Delta\hat{L}_{y}^{2}\right\rangle _{mn} & =e^{-2\xi_{mn}}\sqrt{\left\langle \Delta\hat{L}_{y}^{2}\right\rangle _{mn}\left\langle \Delta\hat{L}_{z}^{2}\right\rangle _{mn}},\\
		\left\langle \Delta\hat{L}_{z}^{2}\right\rangle _{mn} & =e^{2\xi_{mn}}\sqrt{\left\langle \Delta\hat{L}_{y}^{2}\right\rangle _{mn}\left\langle \Delta\hat{L}_{z}^{2}\right\rangle _{mn}},
	\end{align}
	with pseudospin squeezing factor $\xi_{mn}$. Squeezing factor $\xi_{mn}$ is explicitly given by
	\begin{align}
		\xi_{mn} & =-\frac{1}{4}\ln\frac{\left\langle \Delta\hat{L}_{y}^{2}\right\rangle _{mn}}{\left\langle \Delta\hat{L}_{z}^{2}\right\rangle _{mn}}. \label{eq:xi_mn}
	\end{align}
	In the main text, we mention that pseudospin squeezing $\xi_{m0}$ for $m\neq0$ and $n=0$ has a finite value $\xi_\text{inst}\approx r_-$ at the instability $2|D_s| = 1-D_r$. This can be shown, by rewriting $\xi_{m0}$ in the following way
	\begin{align}
		\xi_{m0} &  =-\frac{1}{4}\ln\left(\frac{\left(m+\frac{1}{2}\right)\left(e^{-2r_{-}}+e^{-4r_{+}+2r_{-}}\right)-e^{-2r_{+}}}{\left(m+\frac{1}{2}\right)\left(e^{2r_{-}}+e^{-4r_{+}-2r_{-}}\right)-e^{-2r_{+}}}\right).
	\end{align}
	At the instability, the $\psi_+$ mode becomes unstable as quadrature squeezing diverges $r_+\rightarrow\infty$. We therefore use the approximation $\exp(-r_+)\approx 0$, such that pseudospin squeezing at the instability becomes
	\begin{align}
		\xi_{\text{inst}}  \approx-\frac{1}{4}\ln\left(\frac{e^{-2r_{-}}}{e^{2r_{-}}}\right)
	\end{align}
	which is equivalent to $\xi_\text{inst}\approx r_-$.
	
\begin{figure}
		\centering
		\begin{minipage}{0.49\textwidth}
			(a)\\
			\includegraphics[width=\textwidth]{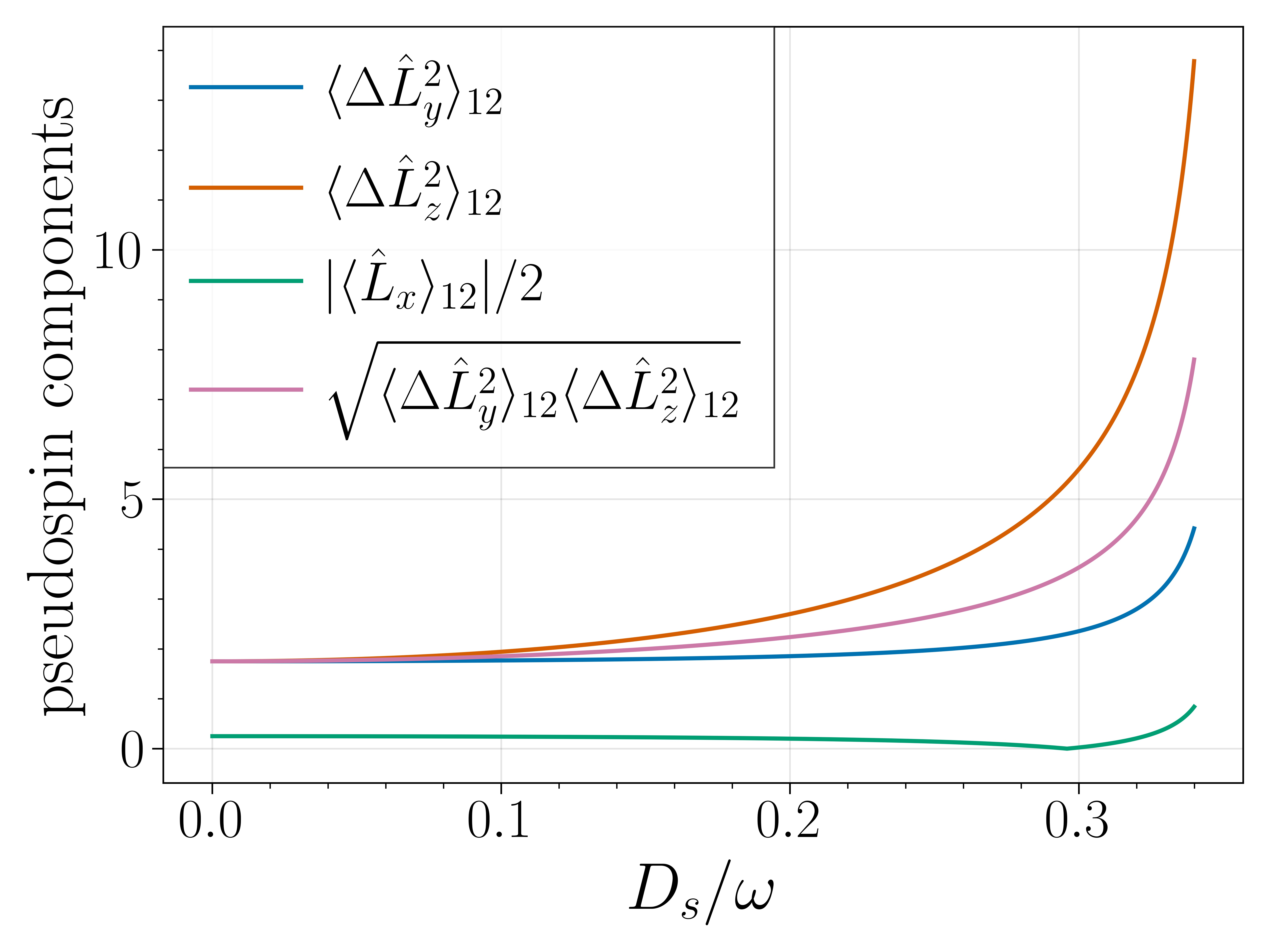}\\
		\end{minipage}
		\begin{minipage}{0.49\textwidth}
			(b)\\
			\includegraphics[width=\textwidth]{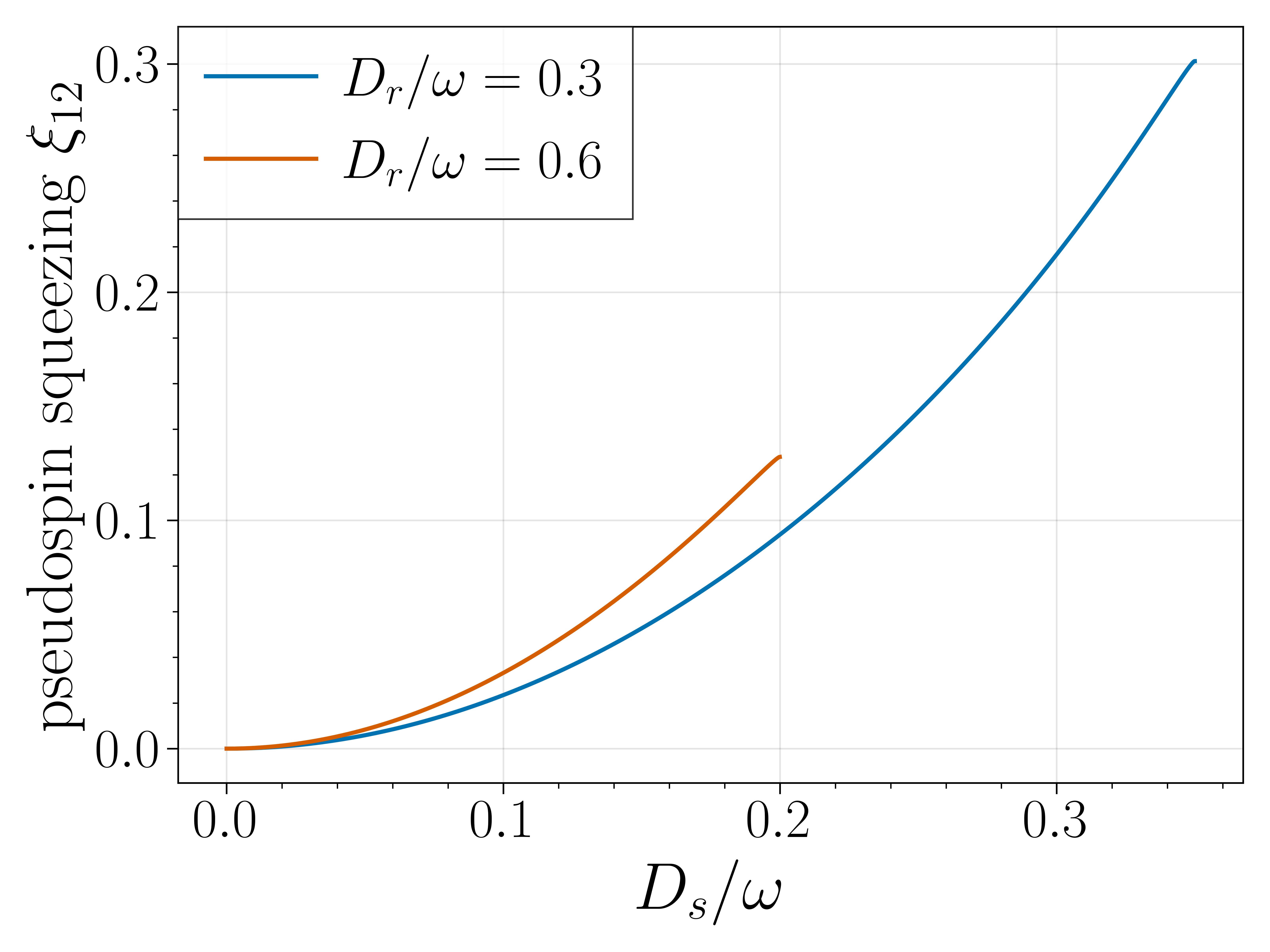}
		\end{minipage}
		\caption{Plot of (a) pseudospin components and variances $|\langle\hat{L}_x\rangle_{12}|$, $\langle\Delta\hat{L}_y^2\rangle_{12}$ and $\langle\Delta\hat{L}_z^2\rangle_{12}$ in the eigenstate $\ket{1,2}$ for fixed $D_r/\omega = 0.3$ (b) pseudsospin squeezing $\xi_{12}$ as a function of $D_s$ . \label{fig:Fig_S1}}
	\end{figure}

	Having discussed squeezing for a state $\ket{m_+,0_-}$, we now want to evaluate Heisenberg's uncertainty relation graphically for states with $m\neq0$ and $n\neq0$ in Fig.~\ref{fig:Fig_S1}(a). One can see for $m=1$ and $n=2$, \al{on one hand $\left|\left\langle \hat{L}_{x}\right\rangle _{12}\right|/2 < \left\langle \Delta\hat{L}_{y}^{2}\right\rangle _{12}$ and therefore fluctuations are larger than the ground state fluctuations. On the other hand, the variances fulfil $\sqrt{\left\langle \Delta\hat{L}_{y}^{2}\right\rangle _{12}} <\sqrt{\left\langle \Delta\hat{L}_{z}^{2}\right\rangle _{12}}$ and hence follows squeezing according to the definition via ellipticity.}
	
	\subsection{Pseudospin squeezing in the ground state}
	In this subsection, we discuss the special case of pseudospin squeezing in the ground state since it differs qualitatively from the previously known case of spin squeezing in its ground state $\ket{0_+,0_-}$ [Eq.~\eqref{eq:Fock_without_qubit}]. Considering $m=0$ and $n=0$ for pseudospin squeezing $\xi_{mn}$ [Eq.~\eqref{eq:xi_mn}], we can already see that $\xi_{00}$ is decreasing with increasing $D_s$. This is unconventional since $D_s$ is what is causing squeezing. The limit $D_s\rightarrow0$ therefore has to be treated carefully in case of the ground state. \al{
		One reason is that the operator $\hat{\boldsymbol{L}}^2 = \left(\hat{N}/2\right)\left(\hat{N}/2+1\right)$~\cite{dalton2017}, with the number operator $\hat{N} = \hat{\alpha}^\dagger\hat{\alpha} + \hat{\beta}^\dagger\hat{\beta}$ has a vanishing expectation value for the magnon vacuum. Hence, in the case where we consider the ground state and the coupling $D_s=0$, the magnitude of pseudospin $\langle\hat{\boldsymbol{L}}^2\rangle_{00}$ vanishes. Therefore, our Bloch sphere has zero radius and our basis is ill-defined. This is the difference between electron spin, where $S$ is always finite, and bosonic pseudospin.
		
		Let's carefully evaluate the limit $D_s\rightarrow 0$} by expanding Eqs.~\eqref{eq:Lx}--\eqref{eq:varLz} for $m=0$ in the small $|D_s|\ll (\omega-D_r)/2$ limit. We find  
	\begin{align}
		\left\langle\hat{L}_x \right\rangle_{00} &\approx \frac{1}{2} \left( \frac{1}{(\omega-D_r)^2} - \frac{1}{(\omega+D_r)^2}\right)|D_s|^2,\\
		\left\langle\Delta\hat{L}_y^2 \right\rangle_{00} &\approx \frac{1}{4} \left( \frac{1}{\omega-D_r} - \frac{1}{\omega+D_r}\right)^2|D_s|^2,\\
		\left\langle\Delta\hat{L}_y^2 \right\rangle_{00} &\approx \frac{1}{4} \left( \frac{1}{\omega-D_r} + \frac{1}{\omega+D_r}\right)^2|D_s|^2.
	\end{align}
	It follows that $\xi_{00}$ in the small $|D_s|\ll (\omega-D_r)/2$ limit has a finite value
	\begin{align}
		\xi &\approx \frac{1}{2}\ln(\frac{\omega}{D_r}).
	\end{align}
	However, the $D_s =0$ limit is ill defined since $\langle\hat{L}_x\rangle_{00} = \langle\Delta\hat{L}_y^2\rangle_{00} = \langle\Delta\hat{L}_z^2\rangle_{00} = 0$. This is because it requires the presence of a nonzero $D_s\neq0$ in order to break symmetry and therefore induce squeezing. The intrasublattice coupling $D_s$ is hence causing squeezing but demonstrates unconventional behaviour in the ground state $\ket{0_+,0_-}$ [Eq.~\eqref{eq:Fock_without_qubit}].
	\begin{figure}
		\centering
		\includegraphics[width=0.5\textwidth]{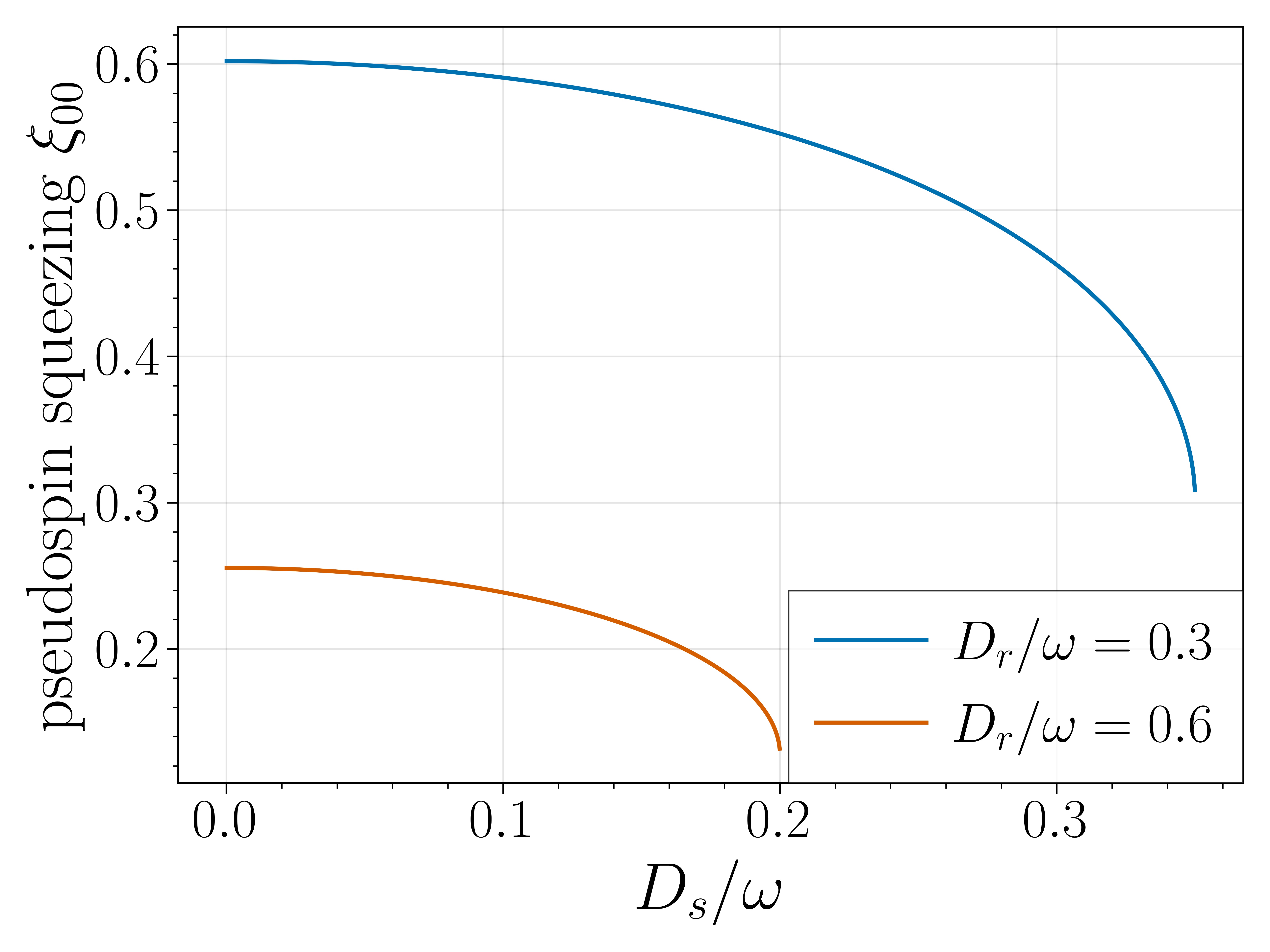}
		\caption{Plot of pseudsospin squeezing $\xi_{00}$ in the ground state $\ket{0,0}$ as a function of $D_s$ .\label{fig:Fig_S2}}
	\end{figure}
	
	\section{Coupling to spin qubit\label{sec:qubit}}
	In this section, we analyze the interaction between an antiferromagnet with strong intrasublattice coupling as described by $\hat{\mathcal{H}}_\mathrm{AFM}$ [Eq.~\eqref{eq:H_AFM_sq}] and a spin qubit in the dispersive regime. In Sec.~\ref{sec:Dis_coup}, we derive the direct dispersive coupling from a spin Hamiltonian. In Sec.~\ref{sec:qubit_diag}, we diagonalize the resulting Hamiltonian describing the AFM, qubit and direct dispersive coupling and discuss state control via the qubit. Finally, in Sec.~\ref{sec:Gs_composition} we discuss if pseudospin squeezing can be resolved via qubit spectroscopy. Here we focus on the signatures of pseudospin squeezing in two coupling regimes.
	\subsection{Dispersive coupling\label{sec:Dis_coup}}
	In this subsection, we derive the direct dispersive coupling between the AFM and the spin qubit. The Hamiltonian describing the spin qubit is given by $\hat{\mathcal{H}}_q = \omega_q\hat{\sigma}_z/2$, where $\omega_q$ denotes the level splitting of the qubit and $\hat{\sigma}_z$ the third Pauli matrix. We assume that the AFM and qubit interact via exchange interaction between the spin qubit and the spins belonging to the AFM interfacing with the qubit~\cite{skogvoll2021,romling2023,kamra2017a}. This interaction can be described by the following Hamiltonian
	\begin{align}
		\hat{\mathcal{H}}_{\text{int}} & =J_{\text{int},A}\sum_{l\in A}\hat{\boldsymbol{S}}_{l}\cdot\hat{\boldsymbol{s}}_{l}+J_{\text{int},B}\sum_{m\in B}\hat{\boldsymbol{S}}_{m}\cdot\hat{\boldsymbol{s}}_{m}\label{eq:H_int}
	\end{align}
	where $l$ and $m$ label the interfacial site belonging to sublattices
	$A$ and $B$, $J_{\text{int},A(B)}$ is the interfacial coupling strength
	\cite{bender2015,kamra2016a,takahashi2010,zhang2012} with lattice site belonging to sublattice $A(B)$, $\hat{\boldsymbol{S}}$
	denotes the spin operator and $\hat{\boldsymbol{s}}$ spin of the
	electronic states comprising the qubit. As argued in~\cite{skogvoll2021} and~\cite{romling2023,romling2024}, terms $\propto \hat{S}_i^x \hat{s}_i^x$ and $\propto \hat{S}_i^y \hat{s}_i^y$ result in a coherent interaction, leading to a hybridization between magnons and the qubit. Here, we want to suppress coherent exchange by choosing a large detuning between the magnon resonance frequency $\omega$ and the qubit level splitting $\omega_q$ and hence focus on the terms $\propto \hat{S}_i^z \hat{s}_i^z$. Using Holstein-Primakoff transformations [Eq.~\eqref{eq:HP}], we find that the terms $\propto \hat{S}_i^z \hat{s}_i^z$ in the interaction Hamiltonian $\hat{\mathcal{H}}_\mathrm{int}$ [Eq.~\eqref{eq:H_int}] have one constant term renormalizing the qubit level splitting~\cite{skogvoll2021} and a coupling term resulting in dispersive interaction~\cite{romling2023}
	\begin{align}
		\hat{\mathcal{H}}_{\text{dis}} & =\left|\phi\right|^{2}\left(\frac{J_{\text{int},A}N_{\text{int},A}}{2N}\hat{a}^{\dagger}\hat{a}-\frac{J_{\text{int},B}N_{\text{int},B}}{2N}\hat{b}^{\dagger}\hat{b}\right)\hat{\sigma}_{z},\label{eq:H_dis}
	\end{align}
	where $|\phi_l|$ denotes the qubit wave function at lattice site $l$, which here is assumed to be homogeneous~\cite{skogvoll2021,romling2023}, and $N_{\mathrm{int},A(B)}$ the number of interfacial lattice sites belonging to sublattice $A(B)$.  
	
	Defining the direct dispersive coupling strengths
	\begin{align}
		\begin{split}
			\chi_{a} & =\frac{J_{\text{int},A}\left|\phi\right|^{2}}{2N}N_{\text{int},A},
		\end{split}
		\begin{split}
			\chi_{b} & =\frac{J_{\text{int},B}\left|\phi\right|^{2}}{2N}N_{\text{int},B},\label{eq:chi_ab}
		\end{split}
	\end{align}
	we can deduce that the coupling between the AFM and qubit depends on the size of the magnet and the structure of the interface. Assuming a compensated magnet with a compensated interface, such that $\chi_a = \chi_b\equiv\chi$, and using the Bogoliubov transformation [Eq.~\eqref{eq:Bogoliubov}], the direct dispersive interaction Hamiltonian $\hat{\mathcal{H}}_\mathrm{dis}$ [Eq.~\eqref{eq:H_dis}] acquires the form used in the main text $\hat{\mathcal{H}}_\mathrm{dis} = 2\chi\hat{L}_z\hat{\sigma}_z$.
	\subsection{Eigenmodes of the coupled AFM and qubit system\label{sec:qubit_diag}}
	In this subsection, we diagonalize the Hamiltonian describing an AFM with strong intrasublattice interaction coupled to a spin qubit via direct dispersive coupling. The full Hamiltonian is given by a sum of the AFM Hamiltonian $\hat{\mathcal{H}}_\mathrm{AFM}$ [Eq.~\eqref{eq:H_AFM_sq}], $\hat{\mathcal{H}}_q = \omega_q\hat{\sigma}_z/2$ and the direct dispersive interaction $\hat{\mathcal{H}}_\mathrm{dis}$ [Eq.~\eqref{eq:H_dis}]. Here, we assume $\chi_a=\chi_b\equiv\chi$ [Eq.~\eqref{eq:chi_ab}], such that the full Hamiltonian in spin-up and spin-down magnon basis reads
	\begin{align}
		\hat{\mathcal{H}}_{0} & =\hat{\mathcal{H}}_\mathrm{AFM}+\frac{\omega_{q}}{2}\hat{\sigma}_{z}+\chi\left(\hat{\alpha}^{\dagger}\hat{\alpha}-\hat{\beta}^{\dagger}\hat{\beta}\right)\hat{\sigma}_{z},\label{eq:H_0}
	\end{align}
	with $\hat{\mathcal{H}}_\mathrm{AFM}$ from Eq.~\eqref{eq:H_AFM_sq}. We start the analysis by projecting the full Hamiltonian $\hat{\mathcal{H}}_0$ [Eq.~\eqref{eq:H_0}] onto the qubit eigenstates, which are given by the ground state $\ket{g}$
	and excited state $\ket{e}$. We define the reduced Hamiltonians by
	\begin{align}
		\begin{split}
			\hat{\mathcal{H}}_{g} & =\bra{g}\hat{\mathcal{H}}_{0}\ket{g},\end{split}
		\begin{split}
			\hat{\mathcal{H}}_{e} & =\bra{e}\hat{\mathcal{H}}_{0}\ket{e}.\label{eq:H_ge}
		\end{split}
	\end{align}
	In the following, we treat $\hat{\mathcal{H}}_{g}$ [Eq.~\eqref{eq:H_ge}]
	in detail and afterwards transfer our results to $\hat{\mathcal{H}}_{e}$
	[Eq.~\eqref{eq:H_ge}] via the substitutions $-\omega_{q}\rightarrow+\omega_{q}$
	and $-\chi\rightarrow+\chi$. We diagonalize $\hat{\mathcal{H}}_g$ using the pseudospin framework with pseudospin operators $\hat{L}_i$, $i\in 0,1,2,3$ [Eq.~\eqref{eq:pseudospin_sm}]. Following~\cite{kamra2020a}, we separate $\hat{\mathcal{H}}_{g}$ [Eq.~\eqref{eq:H_ge}] in a base Hamiltonian and perturbation Hamiltonian, such that $\hat{\mathcal{H}}_{g}=\hat{\mathcal{H}}_{\text{base}}^{g}+\hat{\mathcal{H}}_{\text{pert}}^{g}$
	with 
	\begin{align}
		\hat{\mathcal{H}}_{\text{base}}^{g} & =\omega\left(\hat{\alpha}^{\dagger}\hat{\alpha}+\hat{\beta}^{\dagger}\hat{\beta}\right)+\frac{2\omega-2A-\omega_{q}}{2},\label{eq:H_g_base}\\
		\hat{\mathcal{H}}_{\text{pert,hyb}}^{g} & =-\chi\left(\hat{\alpha}^{\dagger}\hat{\alpha}-\hat{\beta}^{\dagger}\hat{\beta}\right)-D_{r}\left(\hat{\alpha}\hat{\beta}^{\dagger}+\hat{\alpha}^{\dagger}\hat{\beta}\right),\label{eq:H_g_pert_hyb}\\
		\hat{\mathcal{H}}_{\text{pert,quad}}^{g} & =D_{s}\left(\hat{\alpha}^{2}+\hat{\alpha}^{\dagger2}+\hat{\beta}^{2}+\hat{\beta}^{\dagger2}\right).\label{eq:H_g_pert_quad}
	\end{align}
	Here, we divided the perturbation into two parts, the hybridization Hamiltonian $\hat{\mathcal{H}}_{\text{pert,hyb}}^{g}$ [Eq.~\eqref{eq:H_g_pert_hyb}] and quadratic terms $\hat{\mathcal{H}}_{\text{pert,quad}}^{g}$ [[Eq.~\eqref{eq:H_g_pert_quad}]. We begin by treating $\hat{\mathcal{H}}_{\text{base}}^{g} + \hat{\mathcal{H}}_{\text{pert,hyb}}^{g}$ in the pseudospin framework \citep{kamra2020a}, reformulating $\hat{\mathcal{H}}_{\text{base}}^{g} + \hat{\mathcal{H}}_{\text{pert,hyb}}^{g}$ with pseudospin operators $\hat{L}_i$, $i\in 0,1,2,3$ and obtain 
	\begin{align}
		\hat{\mathcal{H}}_{\text{base}}^{g}+\hat{\mathcal{H}}_{\text{pert,hyb}}^{g} & =2\omega\hat{L}_{0}-\boldsymbol{\omega}^{g}\cdot\hat{\boldsymbol{L}}+\frac{2\omega-2A-\omega_{q}}{2}.\label{eq:H_g_base_pert_hyb}
	\end{align}
	with pseudofield 
	\begin{align}
		\boldsymbol{\omega}^{g} & = 2D_{r} \hat{\boldsymbol{e}}_x + 2\chi \hat{\boldsymbol{e}}_z.\label{eq:pf_g}
	\end{align}
	Here, we can already notice that pseudofield $\boldsymbol{\omega}^g$ [Eq.~\eqref{eq:pf_g}] acquires a $z$-component from the interaction with the spin qubit in comparison with $\boldsymbol{\omega}^0$ (with $\chi=0$). As discussed in Sec.~\ref{sec:diagonalization}, the eigenmodes of a Hamiltonian of the form of $\hat{\mathcal{H}}_{\text{base}}^{g} + \hat{\mathcal{H}}_{\text{pert,hyb}}^{g}$ [Eq.~\eqref{eq:H_g_base_pert_hyb}] can be conveniently characterized by the pseudofield $\boldsymbol{\omega}^g$ [Eq.~\eqref{eq:pf_g}] and a rotation operation [Eq.~\eqref{eq:rotation_sa}]. The eigenmodes of $\hat{\mathcal{H}}_{\text{base}}^{g} + \hat{\mathcal{H}}_{\text{pert,hyb}}^{g}$ [Eq.~\eqref{eq:H_g_base_pert_hyb}] can therefore be expressed as
	\begin{align}
		\begin{split}
			\hat{\psi}_1^g & =\hat{R}_{y}\!\left(\theta_g\right)\hat{\alpha}\hat{R}_{y}^{\dagger}\!\left(\theta_g\right),\\
		\end{split}
		\begin{split}
			\hat{\psi}_2^g & =\hat{R}_{y}\!\left(\theta_g\right)\hat{\beta}\hat{R}_{y}^{\dagger}\!\left(\theta_g\right). \label{eq:pseudospin_modes}
		\end{split}
	\end{align}
	with rotation operator $\hat{R}_y\!\left(\theta\right)$ [Eq.~\eqref{eq:R_y_theta}] and rotation angle $\sin\theta_{g}=1/\sqrt{1+\chi^{2}/D_{r}^{2}}$.
	It will be convenient to define a small angle $\tilde{\theta}$ such
	that $\theta_{g}=\frac{\pi}{2}-\tilde{\theta}$, which explicitly reads 
	\begin{align}
		\sin\tilde{\theta} & =|\chi|/\sqrt{\chi^{2}+D_{r}^{2}}.\label{eq:angle_g}
	\end{align}
	Transforming the full Hamiltonian $\hat{\mathcal{H}}_{g}$ [Eq.~\eqref{eq:H_ge}] into the basis of $\hat{\psi}_{1}^{g}$ and $\hat{\psi}_{2}^{g}$ [Eq.~\eqref{eq:pseudospin_modes}], we obtain the expression 
	\begin{align}
		\hat{\mathcal{H}}_{g} & =\omega_{1}^{g}\hat{\psi}_{1}^{g\dagger}\hat{\psi}_{1}^{g}+\omega_{2}^{g}\hat{\psi}_{2}^{g\dagger}\hat{\psi}_{2}^{g} +D_{s}\left(\hat{\psi}_{1}^{g2}+\hat{\psi}_{1}^{g\dagger2}+\hat{\psi}_{2}^{g2}+\hat{\psi}_{2}^{g\dagger2}\right) +\frac{2\omega-2A-\omega_{q}}{2},\label{eq:H_g_psi}
	\end{align}
	with frequencies 
	\begin{align}
		\begin{split}
			\omega_{1}^{g} & =\omega-\sqrt{\chi^{2}+D_{r}^{2}},
		\end{split}
		\begin{split}
			\omega_{2}^{g} & =\omega+\sqrt{\chi^{2}+D_{r}^{2}}.
		\end{split}
	\end{align}
	Now $\hat{\psi}_{1}^{g}$ and $\hat{\psi}_{2}^{g}$ decouple and we can solve $\hat{\mathcal{H}}_g$ [Eq.~\eqref{eq:H_g_psi}] by applying a one-mode squeezing transformation. We end up with the diagonalized Hamiltonian 
	\begin{align}
		\hat{\mathcal{H}}_g & =\omega_{1}^{g,\text{sq}}\hat{\psi}_{1}^{g,\text{sq}\dagger}\hat{\psi}_{1}^{g,\text{sq}}+\omega_{2}^{g,\text{sq}}\hat{\psi}_{2}^{g,\text{sq}\dagger}\hat{\psi}_{2}^{g,\text{sq}}+\frac{\omega_{1}^{g,\text{sq}}+\omega_{2}^{g,\text{sq}}-2A-\omega_{q}}{2},\label{eq:H_g_diag}
	\end{align}
	where the eigenmodes $\hat{\psi}_{1/2}^{g,\mathrm{sq}}$ can be expressed as
	\begin{align}
		\begin{split}
			\hat{\psi}_{1}^{g,\text{sq}} & =\hat{S}_{g,1}^\mathrm{oms}\!\left(r_1^g\right)\hat{\psi}_{1}^{g}\hat{S}_{g,1}^{\mathrm{oms}}\!\left(r_1^g\right)^\dagger,
		\end{split}
		\begin{split}
			\hat{\psi}_{2}^{g,\text{sq}} & =\hat{S}_{g,2}^\mathrm{oms}\!\left(r_2^g\right)\hat{\psi}_{2}^{g}\hat{S}_{g,2}^{\mathrm{oms}}\!\left(r_2^g\right)^\dagger,
		\end{split} \label{eq:psi12-g-sq}
	\end{align}
	with squeeze operators
	\begin{equation}
		\hat{S}_{g,1/2}^{\text{oms}}\!\left(r_{1/2}^g\right) =\exp( \frac{\tilde{r}}{2}\left[\left(\hat{\psi}_{1/2}^{g}\right)^{2}-\left(\hat{\psi}_{1/2}^{g\dagger}\right)^{2}\right]) ,\label{eq:squeeze_g}
	\end{equation}
	and one mode-squeezing factors 
	\begin{align}
		\begin{split}
			\tanh(2r_{1}^{g}) & =\frac{2\left|D_{s}\right|}{\omega-\sqrt{\chi^{2}+D_{r}^{2}}},
		\end{split}
		\begin{split}
			\tanh(2r_{2}^{g}) & =\frac{2\left|D_{s}\right|}{\omega+\sqrt{\chi^{2}+D_{r}^{2}}}.\label{eq:r12}
		\end{split}
	\end{align}
	The eigenenergies in $\hat{\mathcal{H}}_g$ [Eq.~\eqref{eq:H_g_diag}] read
	\begin{align}
		\omega_{1}^{g,\text{sq}} & =\sqrt{\omega^{2}-2\omega\sqrt{\chi^{2}+D_{r}^{2}}+\chi^{2}-4D^{2}},\label{eq:wg_1_sq}\\
		\omega_{2}^{g,\text{sq}} & =\sqrt{\omega^{2}+2\omega\sqrt{\chi^{2}+D_{r}^{2}}+\chi^{2}-4D^{2}}.\label{eq:wg_2_sq}
	\end{align}
	\al{Combining Eqs.~\eqref{eq:pseudospin_modes} and \eqref{eq:psi12-g-sq} allows us to write the eigenmodes of $\hat{\mathcal{H}}_g$ [Eq.~\eqref{eq:H_g_diag}] in the following compact form
		\begin{align}
			\hat{\psi}_{1}^{g,\text{sq}} & =\hat{S}_{g,1}^{\text{oms}}\!\left(r_{1}^g\right)\hat{R}_{y}\!\left(\frac{\pi}{2}-\tilde{\theta}\right)\hat{\alpha}\hat{R}_{y}^{\dagger}\!\left(\frac{\pi}{2}-\tilde{\theta}\right)\hat{S}_{g,1}^{\text{oms}}\!\left(r_{1}^g\right)^{\dagger},\label{eq:psi1_g_sq}\\
			\hat{\psi}_{2}^{g,\text{sq}} & =\hat{S}_{g,2}^{\text{oms}}\!\left(r_{2}^g\right)\hat{R}_{y}\!\left(\frac{\pi}{2}-\tilde{\theta}\right)\hat{\beta}\hat{R}_{y}^{\dagger}\!\left(\frac{\pi}{2}-\tilde{\theta}\right)\hat{S}_{g,2}^{\text{oms}}\!\left(r_{2}^g\right)^{\dagger},\label{eq:psi2_g_sq}
		\end{align}
		which is an equivalent form as Eqs.~\eqref{eq:psi_+} and \eqref{eq:psi_-} from Sec.~\ref{sec:diagonalization}.}
	
	Now we evaluate the diagonalization of the excited state Hamiltonian $\hat{\mathcal{H}}_{e}$
	[Eq.~\eqref{eq:H_ge}] that can be obtained from the solution of $\hat{\mathcal{H}}_{g}$
	[Eq.~\eqref{eq:H_g_diag}] via the substitutions $-\omega_{q}\rightarrow+\omega_{q}$
	and $-\chi\rightarrow+\chi$. Note that we also perform an index substitution $g\rightarrow e$ to indicate entities belonging to the excited state manifold. We find the angle $\theta_e = \pi/2 - \tilde{\theta}$ [Eq.~\eqref{eq:angle_g}], the squeeze factors $r_{1}^{e}=r_{1}^{g}\equiv r_{1}$
	and $r_{2}^{e}=r_{2}^{g}\equiv r_{2}$ [Eq.~\eqref{eq:r12}] and the eigenergies $\omega_{1}^{e,\text{sq}}=\omega_{1}^{g,\text{sq}}\equiv\omega_{1}^{\text{sq}}$ and $\omega_{2}^{e,\text{sq}}=\omega_{2}^{g,\text{sq}}\equiv\omega_{2}^{\text{sq}}$ [Eqs.~\eqref{eq:wg_1_sq} and \eqref{eq:wg_2_sq}]. The diagonalized Hamiltonian $\hat{\mathcal{H}}_{e}$ [Eq.~\eqref{eq:H_ge}] reads 
	\begin{align}
		\hat{\mathcal{H}}_{e} & =\omega_{1}^{\text{sq}}\hat{\psi}_{1}^{e,\text{sq}\dagger}\hat{\psi}_{1}^{e,\text{sq}}+\omega_{2}^{\text{sq}}\hat{\psi}_{2}^{e,\text{sq}\dagger}\hat{\psi}_{2}^{e,\text{sq}}+\frac{\omega_{1}^{\text{sq}}+\omega_{2}^{\text{sq}}-2A+\omega_{q}}{2}.\label{eq:H_e_diag}
	\end{align}
	\al{Equivalently to the expressions for $\hat{\psi}_{1/2}^{q,\mathrm{sq}}$ in Eqs.~\eqref{eq:psi1_g_sq} and \eqref{eq:psi2_g_sq}, }the eigenmodes $\hat{\psi}_{1/2}^{e,\text{sq}}$
	can be obtained by applying the following operations
	\begin{align}
		\hat{\psi}_{1}^{e,\text{sq}} & =\hat{S}_{e,1}^{\text{oms}}\!\left(r_{1}\right)\hat{R}_{y}\!\left(\frac{\pi}{2}+\tilde{\theta}\right)\hat{\alpha}\hat{R}_{y}^{\dagger}\!\left(\frac{\pi}{2}+\tilde{\theta}\right)\hat{S}_{e,1}^{\text{oms}}\!\left(r_{1}\right)^{\dagger},\\
		\hat{\psi}_{2}^{e,\text{sq}} & =\hat{S}_{e,2}^{\text{oms}}\!\left(r_{2}\right)\hat{R}_{y}\!\left(\frac{\pi}{2}+\tilde{\theta}\right)\hat{\beta}\hat{R}_{y}^{\dagger}\!\left(\frac{\pi}{2}+\tilde{\theta}\right)\hat{S}_{e,2}^{\text{oms}}\!\left(r_{2}\right)^{\dagger},
	\end{align}
	with the one-mode squeeze operators 
	\begin{align}
		\hat{S}_{e,1/2}^{\text{oms}}\!\left(r_{1/2}\right) & =\exp( \frac{r_{1/2}}{2}\left[\left(\hat{\psi}_{1/2}^{e}\right)^{2}-\left(\hat{\psi}_{1/2}^{e\dagger}\right)^{2}\right]) .\label{eq:squeeze_e}
	\end{align}
	\al{Using a similar technique and proof as for Eq.~\eqref{eq:Fock_without_qubit} in Sec.~\ref{sec:diagonalization}, we can determine the eigenstates of $\hat{\mathcal{H}}_{g}$ and $\hat{\mathcal{H}}_{e}$ [Eqs.~\eqref{eq:H_g_diag} and \eqref{eq:H_e_diag}]. We find that they can be expressed as}
	\begin{align}
		\ket{m,n}_{g} & =\hat{S}_{g,1}^{\text{oms}}\!\left(r_{1}\right)\hat{S}_{g,2}^{\text{oms}}\!\left(r_{2}\right)\hat{R}_{y}\!\left(\frac{\pi}{2}-\tilde{\theta}\right)\ket{m_{\alpha},n_{\beta}},\label{eq:Fock_g}\\
		\ket{m,n}_{e} & =\hat{S}_{e,1}^{\text{oms}}\!\left(r_{1}\right)\hat{S}_{e,2}^{\text{oms}}\!\left(r_{2}\right)\hat{R}_{y}\!\left(\frac{\pi}{2}+\tilde{\theta}\right)\ket{m_{\alpha},n_{\beta}},\label{eq:Fock_e}
	\end{align}
	where $\ket{m,n}_{g/e}$ denotes a Fock state of $\hat{\mathcal{H}}_{g/e}$
	[Eqs.~\eqref{eq:H_g_diag} and \eqref{eq:H_e_diag}] and $\ket{m_{\alpha},n_{\beta}}$ an eigenstate of $\omega\left(\hat{\alpha}^\dagger\hat{\alpha} + \hat{\beta}^\dagger\hat{\beta}\right)$. We can conclude from this result that the nature of eigenmodes stays the same as in the analysis in Sec.~\ref{sec:diagonalization} but the interaction with a qubit modifies the rotation angle of pseudofield as well as the one-mode squeezing of the eigenmodes. The fact that the rotation angle depends on the state of the qubit ($\ket{g}$ or $\ket{e}$) shows that a qubit can be used for state control of pseudospin. 
	\subsection{Ground state composition\label{sec:Gs_composition}}
	In this section, we discuss if pseudospin squeezing as defined in Sec.~\ref{sec:pseudospin_squeezing} manifests itself via a characteristic signature in qubit spectroscopy~\cite{schuster2007,lachance-quirion2017,kono2017}. For this purpose, we follow the method in~\cite{romling2023,romling2024} and investigate if there is a nonvanishing overlap between the ground state $\ket{0,0}_g$ [Eq.~\eqref{eq:Fock_g}] and an excited state $\ket{m,n,}_e$ [Eq.~\ref{eq:Fock_e}]. 
	
	\al{If there is a nonvanishing overlap between the ground state and an excited state $\vphantom{\ket{0}}_e\!\braket{m,n}{0,0}_g \neq0$ and the qubit is driven, e.g. by an external microwave drive, at a frequency that matches the energy difference between the states $\ket{0,0}_g$ and $\ket{m,n}_e$, then the transition $\ket{0,0}_g \rightarrow \ket{m,n}_e$ occurs with a probability of $|\vphantom{\ket{0}}_e\!\braket{m,n}{0,0}_g|^2$.} When performing qubit spectroscopy by driving the qubit in a range of frequencies and measuring its population, the excitation probability $|\vphantom{\ket{0}}_e\!\braket{m,n}{0,0}_g|^2$ should manifest itself in as a nontrivial peak. 
	
	Our goal in the following is to find out if there are excited states $\ket{m,n}_e$ [Eq.~\ref{eq:Fock_e}] that have a non-vanishing overlap with the ground state $\ket{0,0}_g$ [Eq.~\eqref{eq:Fock_g}], hence demonstrating if pseudospin squeezing has a non-trivial signature in qubit spectroscopy. However, since the operators $\hat{\psi}_{1}^{g}$ and $\hat{\psi}_{2}^{g}$ don't commute
	with $\hat{\psi}_{1}^{e\dagger}$ and $\hat{\psi}_{2}^{e\dagger}$, the analytical calculation of the overlap is a nontrivial task. We therefore expand the operators in the two limits $\chi/D_r\ll1$ and $D_r/\chi\ll1$ and calculate the overlap in these limits.
	
	\subsubsection{Small $\chi$ limit}
	We begin our analysis with the small $\chi$ limit. If $\left|\chi\right|\ll\left|D_{r}\right|$,
	the angle $\tilde{\theta}$ [Eq.~\eqref{eq:angle_g}] becomes
	\begin{align}
		\tilde{\theta} & \approx\frac{\chi}{\left|D_{r}\right|},
	\end{align}
	and correspondingly $\cos\tilde{\theta}\approx\text{sng}\!\left(D\right)$
	and $\sin\tilde{\theta}\approx\tilde{\theta}$. Under
	the same condition $\left|\chi\right|\ll\left|D_{r}\right|$, the squeezing factors $r_{1}$ and $r_{2}$ [Eq.~\eqref{eq:r12}] simplify to 
	\begin{align}
		\begin{split}
			r_{1} & \approx r_{+}+\mathcal{O}\left(\chi^{2}\right),
		\end{split}
		\begin{split}
			r_{2} & \approx r_{-}+\mathcal{O}\left(\chi^{2}\right),
		\end{split}
	\end{align}
	with the squeeze factors $r_{+}$ and $r_{-}$ [Eq.~\eqref{eq:rs_ra}]. Since $\tilde{\theta}$ is first order in $\chi$ and therefore a small parameter, we will express the expansion of operators in terms of $\tilde{\theta}$. The rotation operator $\hat{R}_{y}\!\left(\theta\right)$ [Eq.~\eqref{eq:R_y_theta}] becomes 
	\begin{align}
		\hat{R}_{y}\!\left(\pm\tilde{\theta}\right) & \approx\mathbb{I}\pm\frac{\tilde{\theta}}{2}\left(\hat{\psi}_s\hat{\psi}_{a}^{\dagger}-\hat{\psi}_{s}^{\dagger}\hat{\psi}_{a}\right),
	\end{align}
	with hybridzed modes $\hat{\psi}_{s/a}$ [Eq.~\eqref{eq:psi_sa}]. We note that the one-mode squeezing operators [Eqs.~\eqref{eq:squeeze_g} and \eqref{eq:squeeze_e}] are related via the following relations
	\begin{align}
		\hat{S}_{g,1}^{\text{oms}}\!\left(r_{+}\right) & =\hat{R}_{y}\!\left(-2\tilde{\theta}\right)\hat{S}_{e,1}^{\text{oms}}\!\left(r_{+}\right)\hat{R}_{y}\!\left(2\tilde{\theta}\right),\\
		\hat{S}_{g,2}^{\text{oms}}\!\left(r_{-}\right) & =\hat{R}_{y}\!\left(-2\tilde{\theta}\right)\hat{S}_{e,2}^{\text{oms}}\!\left(r_{-}\right)\hat{R}_{y}\!\left(2\tilde{\theta}\right),
	\end{align}
	which we will use to determine the overlaps between the ground state $\ket{0,0}_{g}$ and excited states
	$\ket{m,n}_{e}$ [Eqs.~\eqref{eq:Fock_g} and \eqref{eq:Fock_e}]. Keeping terms up to first order in $\chi$, we find that the $\ket{0,0}_g$ [Eq.~\eqref{eq:Fock_g}] can be expanded in terms of excited $\ket{m,n}_e$ [Eq.~\eqref{eq:Fock_e}] via the following expression
	\begin{align}
		\ket{0,0}_{g} & \approx\ket{0,0}_{e}+\tilde{\theta}c_{1}\ket{1,1}_{e}
	\end{align}
	with coefficient 
	\begin{align}
		c_{1} & =-\frac{\sinh\left(r_{+}-r_{-}\right)}{2}.
	\end{align}
	From Eqs.~\eqref{eq:H_g_diag} and \eqref{eq:H_e_diag}, we deduce that the energy difference between states $\ket{0,0}_e$ and $\ket{0,0}_g$ is given by $\omega_{00} = \omega_q$ and the energy difference between $\ket{1,1}_e$ and $\ket{0,0}_g$ reads $\omega_{11} = \omega_q + \omega_1^\mathrm{sq}+\omega_2^\mathrm{sq}$ with $\omega_{1/2}^\mathrm{sq}$ from Eqs.~\eqref{eq:wg_1_sq} and \eqref{eq:wg_2_sq}. When driving the qubit at frequency $\omega_{11}$, the transition into state $\ket{1,1}_e$ occurs with a probability of $\theta^2|c_1|^2$. Therefore, qubit spectroscopy would reveal a nontrivial peak around $\omega_{11}$ with a peak height proportional to the transition probability $\theta^2|c_1|^2$. 
	\color{black}
	Considering $D_{s}=0$, we find that the overlap vanishes $c_1 = 0$ which reproduces the result for am AFM with $D_s=0$ with a compensated interface~\cite{romling2024}.
	\subsubsection{Small $D_r$ limit}
	The small $D_r\rightarrow 0$ limit has to be treated separately, since it is dictated by the condition $\left|D_{r}\right|\ll\left|\chi\right|$. The angle $\tilde{\theta}$ [Eq.~\eqref{eq:angle_g}] in the small $D_r$ limit becomes 
	\begin{equation}
		\tilde{\theta}	\approx\frac{\pi}{2}-\frac{\left|D_{r}\right|}{\chi},
	\end{equation}
	and correspondingly $\cos\tilde{\theta}\approx\left|D_{r}\right|/\chi$ and $\sin\tilde{\theta}\approx\text{sgn}\left(D_{r}\right)$. The squeeze factors $r_{1}$ and $r_{2}$ [Eq.~\eqref{eq:r12}] in the small $D_r$ limit read
	\begin{align}
		\begin{split}
			\tanh\left(2r_{1}\right)	&\approx\frac{2\left|D_{s}\right|}{\omega-\left|\chi\right|}+\mathcal{O}\left(D_{r}^{2}\right),
		\end{split}
		\begin{split}
			\tanh\left(2r_{2}\right)	&\approx\frac{2\left|D_{s}\right|}{\omega+\left|\chi\right|}+\mathcal{O}\left(D_{r}^{2}\right).
		\end{split}
	\end{align}
	Here, we consider the zeroth order in $D_{r}$ and therefore take the angle $\tilde{\theta}$ to be $\tilde{\theta}=\pi/2$. We can express the ground state $\ket{0,0}_g$ [Eq.~\eqref{eq:Fock_g}] and excited state $\ket{0,0}_e$ [Eq.~\eqref{eq:Fock_g}]
	\begin{align}
		\begin{split}
			\ket{0,0}_{g}	&=\hat{S}_{\alpha}\!\left(r_{1}\right)\hat{S}_{\beta}\!\left(r_{2}\right)\ket{0_\alpha,0_\alpha},
		\end{split}
		\begin{split}
			\ket{0,0}_{e}	&=\hat{S}_{\beta}\!\left(r_{1}\right)\hat{S}_{\alpha}\!\left(r_{2}\right)\ket{0_\alpha,0_\alpha}, \label{eq:Fock_ge_Dr}
		\end{split}
	\end{align}
	where we used that $\hat{\psi}_{1}^{g}=\hat{\alpha}$, $\hat{\psi}_{2}^{g}=\hat{\beta}$, $\hat{\psi}_{1}^{e}=\hat{\beta}$ and $\hat{\psi}_{2}^{e}=\hat{\alpha}$ if $\tilde{\theta} = \pi/2$. Note that we defined the squeeze operators $\hat{S}_\alpha\!\left(r\right)=\exp(r\left[\hat{\alpha}^2 - \hat{\alpha}^{\dagger2}\right]/2)$ and $\hat{S}_\beta\!\left(r\right)=\exp(r\left[\hat{\beta}^2 - \hat{\beta}^{\dagger2}\right]/2)$. Since $\ket{0,0}_g$ and $\ket{0,0}_e$ [Eq.~\eqref{eq:Fock_ge_Dr}] can be expressed in a common basis, we can connect the two states via the following relation 
	\begin{equation}
		\ket{0,0}_{g}	=\hat{S}_{\alpha}\left(r_{1}-r_{2}\right)\hat{S}_{\beta}\left(r_{2}-r_{1}\right)\ket{0,0}_{e}.
	\end{equation}
	The expansion of the ground state $\ket{0,0}_g$ in terms of excited states $\ket{m,n}_e$ then reads
	\begin{equation}
		\ket{0,0}_g \approx c_{mn}\ket{2m,2n}_e,
	\end{equation}
	with 
	\begin{equation}
		c_{mn} = \frac{(-1)^n}{\cosh(r_\mathrm{eff})}\tanh[m+n](r_\mathrm{eff}) \frac{\sqrt{(2m)!(2n)!}}{2^mm!2^nn!},
	\end{equation}
	and $r_{\text{eff}}=r_{1}-r_{2}$. This result is consistent with~\cite{romling2023} and corresponds to the signature of two decoupled ferromagnets with anisotropies.
	\section{Hematite $\left(\alpha-\text{Fe}_{2}\text{O}_{3}\right)$ and adjustments in formalism}
	In this section, we discuss the antiferromagnet hematite $\left(\alpha-\text{Fe}_{2}\text{O}_{3}\right)$ above the Morin temperature $T_M\approx250\,\mathrm{K}$~\cite{morin1950}, when it is in the canted phase. We first derive the Hamiltonian from a spin model and see that there is a slight modification in comparison with the Hamiltonian discussed in Sec.~\ref{sec:Hamiltonian}. We then discuss how this modification can be incorporated into the theory developed in the main text.
	
	We start with a spin model on a bipartite lattice, consisting of sublattice $A$ and $B$, and consider spin exchange interaction, easy-plane anisotropy and Dzyaloshinskii–Moriya interaction (DMI) and an applied magnetic field along the $y$-axis~\cite{dannegger2023,mazurenko2005,kittel2015,ross2022}
	\begin{align}
		\mathcal{H}_\mathrm{hem}  =J\sum_{\left\langle i,j\right\rangle }&\hat{\boldsymbol{S}}_{i}\cdot\hat{\boldsymbol{S}}_{j}+K\sum_{i\in A,j\in B}\left[\left(\hat{S}_{i}^{x}\right)^{2}+\left(\hat{S}_{j}^{x}\right)^{2}\right]+\mathcal{D}\sum_{\left\langle i,j\right\rangle }\hat{\boldsymbol{x}}\cdot\left[\hat{\boldsymbol{S}}_{i}\times\hat{\boldsymbol{S}}_{j}\right] \nonumber\\
		&-\gamma\mu_{0}H_{0}\sum_{i\in A,j\in B}\left(\hat{S}_{i}^{y}+\hat{S}_{j}^{y}\right)
	\end{align}
	where $J$ denotes the exchange integral, $K$ the anisotropy coefficient, $\mathcal{D}$ the DMI strength and $H_0$ the magnitude of the applied magnetic field. Following~\cite{boventer2023} and~\cite{wimmer2020}, we use the following Holstein-Primakoff and Fourier transformation
	\begin{align}
		\begin{split}
			\hat{S}_{j}^{+} & =\sqrt{\frac{2S}{N}}\sum_{\boldsymbol{k}}e^{-i\boldsymbol{k}\cdot\boldsymbol{r}}\hat{a}_{\boldsymbol{k}}^{\dagger},\\
			\hat{S}_{j}^{-} & =\sqrt{\frac{2S}{N}}\sum_{\boldsymbol{k}}e^{i\boldsymbol{k}\cdot\boldsymbol{r}}\hat{a}_{\boldsymbol{k}},\\
			\hat{S}_{j}^{z^{\prime\prime}} & =-S+\frac{1}{N}\sum_{\boldsymbol{k},\boldsymbol{k}^{\prime}}e^{-i\left(\boldsymbol{k}-\boldsymbol{k}^{\prime}\right)\cdot\boldsymbol{r}}\hat{a}_{\boldsymbol{k}}^{\dagger}\hat{a}_{\boldsymbol{k}^{\prime}}. \label{eq:HP_hematite}
		\end{split}
		\begin{split}
			\hat{S}_{i}^{+} & =\sqrt{\frac{2S}{N}}\sum_{\boldsymbol{k}}e^{i\boldsymbol{k}\cdot\boldsymbol{r}}\hat{b}_{\boldsymbol{k}},\\
			\hat{S}_{i}^{-} & =\sqrt{\frac{2S}{N}}\sum_{\boldsymbol{k}}e^{-i\boldsymbol{k}\cdot\boldsymbol{r}}\hat{b}_{\boldsymbol{k}}^{\dagger},\\
			\hat{S}_{i}^{z^{\prime}} & =S-\frac{1}{N}\sum_{\boldsymbol{k},\boldsymbol{k}^{\prime}}e^{-i\left(\boldsymbol{k}-\boldsymbol{k}^{\prime}\right)\cdot\boldsymbol{r}}\hat{b}_{\boldsymbol{k}}^{\dagger}\hat{b}_{\boldsymbol{k}^{\prime}},
		\end{split}
	\end{align}
	where index $i$($j$) denotes a lattice site belonging to sublattice $A$($B$) and with $\hat{S}_{i}^{\pm}=\hat{S}_{i}^{x}\pm i\hat{S}_{i}^{y^{\prime\prime}}$ and $\hat{S}_{j}^{\pm}=\hat{S}_{j}^{x}\pm\hat{S}_{j}^{y^{\prime}}$. \al{The directions $y^\prime$, $y^{\prime\prime}$, $z^\prime$ and $z^{\prime\prime}$ can be obtained via a rotation around the $x$-axis~\cite{wimmer2020} and are schematically depicted in the canted ground state of hematite in Fig.~\ref{fig:hematite}.
		\begin{figure}
			\centering
			\includegraphics{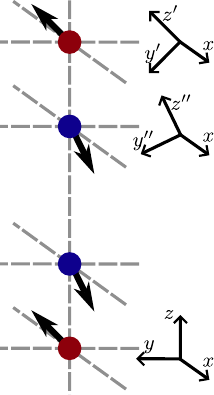}
			\caption{Canted ground state of hematite above Morin temperature $T_M$. The spins of the down spin sublattice (depicted in blue) point along the $-z^{\prime\prime}$ direction, whereas the spin of the up-spin sublattice (depicted in red) point along the $z^\prime$ direction. The primed coordinates are obtained from a rotation of of the coordinte system denoted by $x$, $y$ and $z$ via a rotation around the $x$-axis~\cite{wimmer2020}. \label{fig:hematite}}
		\end{figure}
	}The operators $\hat{a}^\dagger$($\hat{b}^\dagger$) denote the creation operators of spin-up (down) sublattice magnons. Applying this transformation [Eq.~\eqref{eq:HP_hematite}] to the Hamiltonian $\hat{\mathcal{H}}_\mathrm{hem}$ leads to the following expression in terms of sublattice magnons
	\begin{align}
		\hat{\mathcal{H}}_\mathrm{hem}  = \sum_{\boldsymbol{k}}\Big[A&\hat{a}_{\boldsymbol{k}}^{\dagger}\hat{a}_{\boldsymbol{k}}+B\hat{b}_{\boldsymbol{k}}^{\dagger}\hat{b}_{\boldsymbol{k}}+C_{\boldsymbol{k}}\left(\hat{a}_{\boldsymbol{k}}\hat{b}_{-\boldsymbol{k}}+\hat{a}_{\boldsymbol{k}}^{\dagger}\hat{b}_{-\boldsymbol{k}}^{\dagger}\right) \nonumber \\
		&+D\left(\hat{a}_{\boldsymbol{k}}\hat{a}_{-\boldsymbol{k}}+\hat{b}_{\boldsymbol{k}}\hat{b}_{-\boldsymbol{k}}+\hat{a}_{\boldsymbol{k}}^{\dagger}\hat{a}_{-\boldsymbol{k}}^{\dagger}+\hat{b}_{\boldsymbol{k}}^{\dagger}\hat{b}_{-\boldsymbol{k}}^{\dagger}\right)+E_{\boldsymbol{k}}\left(\hat{a}_{\boldsymbol{k}}\hat{b}_{\boldsymbol{k}}^{\dagger}+\hat{a}_{\boldsymbol{k}}^{\dagger}\hat{b}_{\boldsymbol{k}}\right)\Big], \label{eq:hem_mag}
	\end{align}
	with 
	\begin{align}
		\begin{split}
			A & \approx JSz+KS,\\
			B & \approx JSz+KS,,\\
			C_{\boldsymbol{k}} & =JSz\gamma_{\boldsymbol{k}},
		\end{split}
		\begin{split}
			D & =\frac{KS}{2},\\
			E_{\boldsymbol{k}} & =\phi\mathcal{D}Sz\gamma_{\boldsymbol{k}},\\
			\phi & =\frac{-\gamma\mu_{0}\hbar H_{0}-\mathcal{D}Sz}{2JSz},
		\end{split}
	\end{align}
	where $z$ is the number of nearest neighbors.
	
	Finally, assuming a nanomagnet and taking only into account the $\boldsymbol{k}=\boldsymbol{0}$ mode~\cite{skogvoll2021,romling2024}, it follows $\gamma_{\boldsymbol{k}}=1$ and we define $\hat{a}\equiv\hat{a}_{\boldsymbol{0}}$ , $\hat{b}\equiv\hat{b}_{\boldsymbol{0}}$,
	$C\equiv C_{\boldsymbol{0}}$, $E\equiv E_{\boldsymbol{0}}$. Our Hamiltonian describing hematite $\hat{\mathcal{H}}_\mathrm{hem}$ [Eq.~\eqref{eq:hem_mag}] becomes
	\begin{align}
		\hat{\mathcal{H}}_\mathrm{hem} & =A\hat{a}^{\dagger}\hat{a}+B\hat{b}^{\dagger}\hat{b}+C\left(\hat{a}\hat{b}+\hat{a}^{\dagger}\hat{b}^{\dagger}\right)+D\left(\hat{a}^{2}+\hat{b}^{2}+\hat{a}^{\dagger2}+\hat{b}^{\dagger2}\right)+E\left(\hat{a}\hat{b}^{\dagger}+\hat{a}^{\dagger}\hat{b}\right). \label{eq:hem_k0}
	\end{align}
	It is now evident that $\hat{\mathcal{H}}_\mathrm{hem}$ [Eq.~\eqref{eq:hem_k0}] contains extra terms $\hat{a}\hat{b}^{\dagger},\hat{a}^{\dagger}\hat{b}$ that are spin non-conserving in comparison with the NiO Hamiltonian $\hat{H}_\mathrm{NiO}$ [Eq.~\eqref{eq:H_AFM}]. We now want to tackle the question how our previous theory analysis changes when incorporating more spin non-conserving terms $\propto \left(\hat{a}\hat{b}^{\dagger}+\hat{a}^{\dagger}\hat{b}\right)$.
	
	We start by applying the Bogoliubov transformation [Eq.~\eqref{eq:Bogoliubov}] on $\hat{\mathcal{H}}_\mathrm{hem}$ [Eq.~\eqref{eq:hem_k0}], such that
	\begin{align}
		\hat{\mathcal{H}}_\mathrm{hem} & =\omega\left(\hat{\alpha}^{\dagger}\hat{\alpha}+\hat{\beta}^{\dagger}\hat{\beta}\right)-\tilde{D}_r\left(\hat{\alpha}\hat{\beta}^{\dagger}+\hat{\alpha}^{\dagger}\hat{\beta}\right)+\tilde{D}_s\left(\hat{\alpha}^{2}+\hat{\alpha}^{\dagger2}+\hat{\beta}^{2}+\hat{\beta}^{\dagger2}\right)+\left(\omega-A\right).\label{eq:H_hem_sq}
	\end{align}
	with the modified coupling strengths
	\begin{align}
		\begin{split}
			\tilde{D}_r = D\left(\sinh(2r) - \frac{E}{2D}\sinh(2r)\right),
		\end{split}
		\begin{split}
			\tilde{D}_s = D\left(\cosh(2r) - \frac{E}{D}\cosh(2r)\right).
		\end{split}
	\end{align}
	Our analysis from Secs.~\ref{sec:diagonalization}, \ref{sec:pseudospin_squeezing} and \ref{sec:qubit} can therefore be conveniently applied to hematite by substituting $D_r \rightarrow \tilde{D}_r$ and $D_s \rightarrow \tilde{D}_s$.
	
	\section{Rotation transformation for pseudospin \label{sec:Rotation-transformation}}
	\al{In this section, we demonstrate the relations $\psi_{s}=\hat{R}_{y}\left(\theta\right)\hat{\alpha}\hat{R}_{y}^{\dagger}\left(\theta\right)$ and $\psi_{a}=\hat{R}_{y}\left(\theta\right)\hat{\beta}\hat{R}_{y}^{\dagger}\left(\theta\right)$ from Eq.~\eqref{eq:rotation_sa} with the rotation matrix $\hat{R}_{y}\left(\theta\right)  =\exp\left(-i\theta\hat{L}_{y}\right)$ and the pseudospin component $\hat{L}_{y}=\frac{i}{2}\left(\hat{\alpha}\hat{\beta}^{\dagger}-\hat{\alpha}^{\dagger}\hat{\beta}\right)$.
		From the Baker-Campbell-Hausdorff formula~\cite{baker1901} follows
		\begin{align}
			e^{\hat{A}}\hat{\alpha}e^{-\hat{A}} & =\hat{\alpha}+\left[\hat{A},\hat{\alpha}\right]+\frac{1}{2}\left[\hat{A},\left[\hat{A},\hat{\alpha}\right]\right]+\dots\\
			& =\hat{\alpha}+\sum_{n=1}\frac{1}{n!}\left[\hat{A},\hat{\alpha}\right]_{n},\label{eq:BCH-expansion}
		\end{align}
		where $\left[\hat{A},\hat{\alpha}\right]_{n}$ is the $n$-fold nested
		commutator $\left[\hat{A},\hat{\alpha}\right]_{n}=\left[\hat{A},\left[\hat{A},\hat{\alpha}\right]_{n-1}\right]$
		and $\left[\hat{A},\hat{\alpha}\right]_{1}=\left[\hat{A},\hat{\alpha}\right]$. Our rotation operations $\psi_{s}=\hat{R}_{y}\left(\theta\right)\hat{\alpha}\hat{R}_{y}^{\dagger}\left(\theta\right)$ and $\psi_{a}=\hat{R}_{y}\left(\theta\right)\hat{\beta}\hat{R}_{y}^{\dagger}\left(\theta\right)$ from Eq.~\eqref{eq:rotation_sa}
		can be expressed with the expansion expansion in Eq.~\eqref{eq:BCH-expansion}. Let's define for convenience
		$\hat{A}=\frac{\theta}{2}\left(\hat{\alpha}\hat{\beta}^{\dagger}-\hat{\alpha}^{\dagger}\hat{\beta}\right)$
		and determine $\left[\hat{A},\hat{\alpha}\right]_{n}$. We find 
		\begin{align}
			\left[\hat{A},\hat{\alpha}\right]_{n} & =\begin{cases}
				\left(\frac{\theta}{2}\right)^{2k+1}\left(-1\right)^{k}\hat{\beta} & \text{for }n=2k+1,\\
				\left(\frac{\theta}{2}\right)^{2k}\left(-1\right)^{k}\hat{\alpha} & \text{for }n=2k,
			\end{cases}\label{eq:nested-commutator}
		\end{align}
		with $k\in\mathbb{N}$. This can be shown with mathematical induction. Let's start with $n=1$
		\begin{align}
			\left[\hat{A},\hat{\alpha}\right]_{1} & =\frac{\theta}{2}\left[\left(\hat{\alpha}\hat{\beta}^{\dagger}-\hat{\alpha}^{\dagger}\hat{\beta}\right),\hat{\alpha}\right] =\frac{\theta}{2}\hat{\beta},
		\end{align}
		and $n=2$ 
		\begin{align}
			\left[\hat{A},\hat{\alpha}\right]_{2} & =\left(\frac{\theta}{2}\right)^{2}\left[\left(\hat{\alpha}\hat{\beta}^{\dagger}-\hat{\alpha}^{\dagger}\hat{\beta}\right),\hat{\beta}\right] =-\left(\frac{\theta}{2}\right)^{2}\hat{\alpha},
		\end{align}
		which is both in accordance with the formula for the nested commutator in Eq.~\eqref{eq:nested-commutator}. We now perform the induction step for $n+1$, where we have to make a distinction between $n+1$ even and $n+1$ odd. We begin with $n+1 = 2k+1$. The nested commutator for $2k+1$ then reads
		\begin{align}
			\left[\hat{A},\hat{\alpha}\right]_{2k+1} & =\left[\hat{A},\left[\hat{A},\hat{\alpha}\right]_{2k}\right] =\left(\frac{\theta}{2}\right)^{2k}\left(-1\right)^{k}\left[\hat{A},\hat{\alpha}\right]\\
			& =\left(\frac{\theta}{2}\right)^{2k+1}\left(-1\right)^{k}\hat{\beta},
		\end{align}
		which reproduces the formula in Eq.~\eqref{eq:nested-commutator}. Now we take an even $n+1=2k+2$. The corresponding nested commutator reads
		\begin{align}
			\left[\hat{A},\hat{\alpha}\right]_{2k+2} & =\left[\hat{A},\left[\hat{A},\hat{\alpha}\right]_{2k+1}\right] =\left(\frac{\theta}{2}\right)^{2k+1}\left(-1\right)^{k}\left[\hat{A},\hat{\beta}\right]\\
			& =\left(\frac{\theta}{2}\right)^{2k+2}\left(-1\right)^{k+1}\hat{\alpha},
		\end{align}
		which is also in accordance with the formula in Eq.~\eqref{eq:nested-commutator}. We can now separate
		odd and even $n$ in the expansion of $e^{\hat{A}}\hat{\alpha}e^{-\hat{A}}$ [Eq.~\eqref{eq:BCH-expansion}], which becomes
		\begin{align}
			e^{\hat{A}}\hat{\alpha}e^{-\hat{A}} & =\hat{\alpha}+\sum_{k=1}\frac{1}{\left(2k\right)!}\left[\hat{A},\hat{\alpha}\right]_{2k}+\sum_{k=0}\frac{1}{\left(2k+1\right)!}\left[\hat{A},\hat{\alpha}\right]_{2k+1}\\
			& =\sum_{k=0}\frac{1}{\left(2k\right)!}\left(\frac{\theta}{2}\right)^{2k}\left(-1\right)^{k}\hat{\alpha}+\sum_{k=0}\frac{1}{\left(2k+1\right)!}\left(\frac{\theta}{2}\right)^{2k+1}\left(-1\right)^{k}\hat{\beta}\\
			& =\cos\left(\frac{\theta}{2}\right)\hat{\alpha}+\sin\left(\frac{\theta}{2}\right)\hat{\beta}.
		\end{align}
		With this we show 
		\begin{align}
			\hat{R}_{y}\left(\theta\right)\hat{\alpha}\hat{R}_{y}^{\dagger}\left(\theta\right)= & \cos\left(\frac{\theta}{2}\right)\hat{\alpha}+\sin\left(\frac{\theta}{2}\right)\hat{\beta},
		\end{align}
		and therefore 
		\begin{align}
			\psi_{s} & =\hat{R}_{y}\left(\theta\right)\hat{\alpha}\hat{R}_{y}^{\dagger}\left(\theta\right).
		\end{align}
		Under the same considerations, we can show that $\psi_{a}=\hat{R}_{y}\left(\theta\right)\hat{\beta}\hat{R}_{y}^{\dagger}\left(\theta\right).$
		Now the nested commutators become 
		\begin{align*}
			\left[\hat{A},\hat{\beta}\right]_{n} & =\begin{cases}
				\left(\frac{\theta}{2}\right)^{2k+1}\left(-1\right)^{k+1}\hat{\alpha} & \text{for }n=2k+1,\\
				\left(\frac{\theta}{2}\right)^{2k}\left(-1\right)^{k}\hat{\beta} & \text{for }n=2k,
			\end{cases}
		\end{align*}
		such that 
		\begin{align}
			\hat{R}_{y}\left(\theta\right)\hat{\beta}\hat{R}_{y}^{\dagger}\left(\theta\right)= & \cos\left(\frac{\theta}{2}\right)\hat{\beta}-\sin\left(\frac{\theta}{2}\right)\hat{\alpha},
		\end{align}
		from which follows indeed $\psi_{a}=\hat{R}_{y}\left(\theta\right)\hat{\beta}\hat{R}_{y}^{\dagger}\left(\theta\right)$.}
	
\end{document}